\documentclass[reprint,aps,twocolumn,superscriptaddress,nofootinbib]{revtex4-2}

\usepackage{bm}
\usepackage{dcolumn}
\usepackage
[colorlinks=true, allcolors=blue]
{hyperref}

\usepackage{caption} 
\usepackage{ragged2e} 

\newcommand{\ket}[1]{|{#1}\rangle}

\usepackage{graphicx}
\usepackage{tabularx}
\usepackage{nicefrac}
\usepackage{amsmath}
\usepackage{comment}
\usepackage{xcolor}
\begin{document}

\preprint{APS/123-QED}

\title{Fully integrated quantum frequency processor on a silicon chip}%

\author{Sara Congia}
 \email{sara.congia02@universitadipavia.it}
\affiliation{Dipartimento di Fisica "Alessandro Volta", Università di Pavia, via Bassi 6, 27100 Pavia, Italy}
\affiliation{Dipartimento di Ingegneria Industriale e dell’Informazione, Università di Pavia, via Ferrata 1, 27100 Pavia, Italy}
 
\author{Leopold Virot}
\affiliation{STMicroelectronics, Crolles, France}

\author{Elena Rovetta}
\affiliation{Dipartimento di Fisica "Alessandro Volta", Università di Pavia, via Bassi 6, 27100 Pavia, Italy}

\author{Antonio Fincato}
\affiliation{STMicroelectronics, Crolles, France}

\author{Frederic Boeuf}
\affiliation{STMicroelectronics, Crolles, France}

\author{Matteo Galli}
\affiliation{Dipartimento di Fisica "Alessandro Volta", Università di Pavia, via Bassi 6, 27100 Pavia, Italy}

\author{Daniele Bajoni}
\affiliation{Dipartimento di Ingegneria Industriale e dell’Informazione, Università di Pavia, via Ferrata 1, 27100 Pavia, Italy}

\author{Massimo Borghi}
\affiliation{Dipartimento di Fisica "Alessandro Volta", Università di Pavia, via Bassi 6, 27100 Pavia, Italy}

\date{\today}

\begin{abstract}
Frequency-bin encoding has recently emerged as a powerful approach for photonic quantum  information processing, offering high dimensionality, gate-parallelization, and compatibility with existing telecommunication infrastructure. However, its scalable deployment has so far been hindered by the lack of an integrated platform capable of unifying quantum state generation, coherent frequency mixing, and programmable spectral control.\\
Here, we report the first fully integrated quantum frequency processor, monolithically integrating on the same silicon photonic chip a microresonator-based biphoton quantum frequency comb source, a pump-rejection filter, high-speed phase modulators, and a four-channel, line-by-line pulse shaper. We demonstrate key functionalities, such as tunable frequency beamsplitters with success probabilities exceeding $94\%$ and fidelities above $99.9\%$, as well as the ability to synthesize more general single-qubit gates. Finally, we generate and coherently manipulate high-dimensional frequency-bin entangled states entirely on chip, showcasing control over two-photon quantum walks and performing the first on-chip frequency-bin quantum state tomography of a Bell-state with a fidelity of $95.7(3)\%$. By integrating all key functional elements on the same $4\times7\,\textrm{mm}^2$ chip, with the possibility of scaling to a larger number of modes, our work marks an important step toward large-scale frequency-domain photonic processors for both classical and quantum applications.
\end{abstract}

\maketitle

\section*{Introduction}
\label{sec:intro}
Integrated optical processors are chip-scale systems that generate, coherently manipulate, and transmit information using light \cite{shi2020scaling,xie2022low,han2025exploring}.
To date, the most advanced photonic processors for classical data processing—in terms of integration density, reliability, performance, and component count—are capable of flexibly exploiting multiple optical degrees of freedom (DoF). Through large-scale programmable networks of beamsplitters and phase shifters, spectral filters, multimode waveguides, optical delay lines, and high-speed modulators, these systems can indiscriminately manipulate spatial, spectral, temporal, and polarization DoFs, enabling coherent transmission as well as analog optical and microwave signal processing \cite{deng2025single,shi2020scaling,liu2022silicon,wang2015reconfigurable,tao2022fully,xie2022low}.\\
In sharp contrast, large-scale integrated quantum photonic processors predominantly rely on dual-rail or, more generally, multi-rail encodings \cite{bao2023very,qiang2018large,maring2024versatile,vigliar2021error}, a design choice largely driven by the need to manipulate fragile quantum states of light with minimal loss and high operational fidelity using mature linear-optical interferometric networks. While such encoding is well suited for on-chip processing, it is intrinsically inefficient for quantum state distribution and for scaling to high-dimensional Hilbert spaces, as it requires multiple spatial modes. In this context, alternative photonic DoFs—such as time-bin \cite{finco2024time,yu2025quantum}, transverse-mode \cite{feng2022transverse,forbes2025hybrid}, and polarization \cite{zhang2024polarization,miloshevsky2024cmos}—are particularly attractive and objects of active research, as they are naturally compatible with fiber-based transmission and enable high-dimensional encoding within a single spatial mode.\\
Recently, the frequency degree of freedom—traditionally used for multiplexing in both classical and quantum networks—has emerged as a basis for encoding quantum information itself \cite{2016_Lukens_QFP_theory}. In this framework, logical states are defined over discrete, narrowband, and non-overlapping spectral intervals, referred to as frequency bins (FBs)  \cite{2023_Lu_FB_review,2025_Lukens_FBreview,wang2025large}. FBs encoding possess compelling features, including single-spatial-mode operation, very high dimensionality \cite{imany2019high}, gate parallelization \cite{henry2023parallelizable,henry2024parallelization}, and compatibility with existing telecommunication infrastructure \cite{chapman2025quantum}. 
Moreover, this encoding is naturally matched to on-chip quantum light sources based on biphoton frequency combs (BPFC) generated in optical microcavities \cite{2017_Kues_FB}. 
Over the past few years, FB quantum information processing has witnessed a significant progress, driven by both theoretical studies \cite{2016_Lukens_QFP_theory,henry2023parallelizable,fabre2022time,lukens2026paradigm} and experimental demonstrations, including quantum state engineering \cite{2023_Borghi,clementi2023programmable}, tomography \cite{lu2022bayesian}, Hong-Ou-Mandel interference \cite{joshi2020frequency}, metrology \cite{jing2025hong,seshadri2022nonlocal}, quantum walks \cite{2020_Imany_QWalk,2022_Haldar_QWalk}, quantum key distribution \cite{2025_Kues_FB_QKD,2025_Belabas_FB_QKD,2024_Tagliavacche}, and continuous-variable cluster-state generation \cite{jia2025continuous,wang2025large}.\\
An important step toward a complete framework for quantum information processing with FBs was established by Lukens and coworkers \cite{2016_Lukens_QFP_theory}, who introduced the Quantum Frequency Processor (QFP) as a fundamental processing unit combining coherent frequency mixing with line-by-line spectral phase control.\\
Bulk implementations of the QFP have enabled arbitrary single-qubit rotations \cite{2018_Lu_FB_BS}, high-dimensional discrete Fourier transforms \cite{lu2022high}, Bell-state analysis \cite{lingaraju2022bell}, and controlled-NOT gates \cite{lu2019controlled}. Moving toward integrated platforms, electro-optic frequency beam splitters have been implemented in thin-film lithium niobate using coupled resonators \cite{zhu2022spectral}, alongside frequency shifting and bandwidth compression of heralded single photons \cite{assumpcao2024thin}. In parallel, integrated line-by-line pulse shapers based on silicon nitride microresonators have achieved spectral resolutions down to 3 GHz \cite{cohen2024silicon} and have been used to control time correlations of entangled photon pairs \cite{wu2025chip}. 
Fast electro-optic modulation in thin-film lithium niobate has also been heterogeneously integrated with silicon microresonators to generate photon pairs with tunable spectral purity \cite{wang2023integrated}. Coherent FB manipulation has also been shown through all-optical approaches \cite{2025_Oliver_NwayBS,joshi2020frequency,folge2024framework} and very recently theoretically proposed using acousto-optic modulation and multi-modal phase matching in waveguides \cite{lukens2026paradigm}.\\
Despite these advances, frequency-bin quantum information processing remains less mature than some other DoFs. A major limitation towards scalability has been the lack of a fully integrated platform that simultaneously supports nonclassical light generation, high-speed coherent frequency manipulation, and programmable spectral control.\\ 
In this work, we introduce the first fully integrated QFP that unifies all three capabilities on the silicon photonic platform. Our processor integrates a microresonator-based biphoton quantum frequency comb source with controllable linewidth, a pump-rejection filter, two high-speed phase modulators, and a four-channel microring-resonator-based waveshaper. We demonstrate tunable frequency beamsplitters using single-tone RF modulation, achieving success probabilities exceeding $94\%$ and fidelities above $99.9\%$. Following Ref.~\cite{2018_Lu_FB_BS}, we implement general single-qubit gates through RF-delay control and programmable spectral phases.
Finally, we demonstrate full quantum functionality by generating photon pairs entangled across multiple frequency modes via spontaneous four-wave mixing. High dimensional states are coherently manipulated to show both ballistic  and strongly confined energy transport in two-photon quantum walks \cite{2020_Imany_QWalk}. Moreover, two-qubit Bell states are carved from the high-dimensional quantum microcomb, and we implement the first on-chip frequency-bin quantum state tomography (QST) with a fidelity of $95.7(3)\%$.\\

\section*{Results}
\subsection*{Description of the programmable frequency processor}
\label{sec:exp_setup}
\begin{figure*}[p]
\centering
\includegraphics[width = \textwidth]{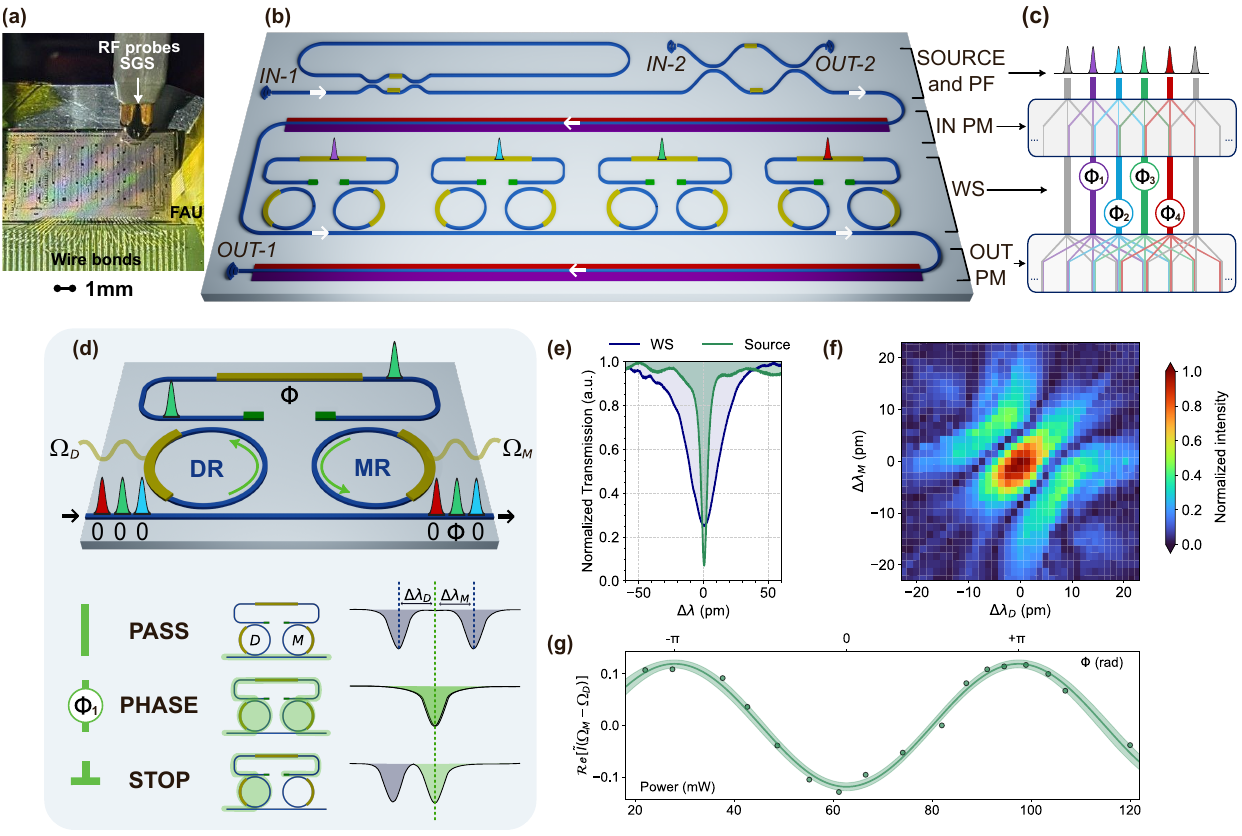} 
\caption{\parbox{\textwidth}{\noindent\justifying
\textbf{Integrated quantum frequency processor.} (a) Photograph of the $4 \times 9 \, \mathrm{mm}^2$ chip, indicating the fiber-array unit (FAU) for optical input–output coupling, the external PCB wire-bonded to the on-chip heater pads (wire bonds), and the RF probes in the signal–ground–signal (SGS) configuration.
(b) Schematic of the integrated silicon photonic quantum frequency processor.
SOURCE: microresonator-based photon-pair source, PF: Mach–Zehnder–based pump-rejection filter, IN PM: input fast phase modulator, WS: four-channel line-by-line waveshaper, OUT PM: output fast phase modulator.
Waveguides are shown in blue, heaters in yellow, Germanium absorbers in green, and the p–n junctions of the modulators are indicated in red and purple. The pump(calibration) light is injected through the IN-1(IN-2) port, while photon pairs(filtered pump) are extracted from the OUT-1(OUT-2) port.
(c) High-level schematic of the processor operation. From top to bottom, the on-chip generated input frequency bins first undergo mode mixing in the IN PM, then acquire programmable spectral phases $\{\Phi_j\}_{j=1}^{4}$ imparted by the four-channel WS, and are finally mixed again by the OUT PM.
(d) Top: dual-ring WS-unit working principle. The DEMUX ring (DR) drops a selected frequency bin (green), which is phase-shifted by $\Phi$ and subsequently multiplexed back into the common bus waveguide by the MUX ring (MR). DC-voltages and separate dither tones $\Omega_{D(M)}$ are applied to the heaters for wavelength tuning and calibration purposes. Bottom: programmable channel operations, each associated with different light paths and relative resonance detuning $\Delta\lambda_{D(R)}$ of the DR(MR) ring with respect to the target channel wavelength (dashed green line).
(e) Spectral response of a WS channel at $\Phi=\pi$ and of one resonances of the resonator source. Both spectra are reported as a function of the wavelength detuning  $\Delta\lambda$ from the center wavelength of the WS channel. 
(f) Measured amplitude of the Fourier transform $\tilde{I}$ at frequency $2(\Omega_M+\Omega_D)$ of the time-dependent signal intensity $I(t)$ at the output of a WS channel during the calibration phase. See Methods section for more details on the calibration procedure.  
(g) Calibration of the phase $\Phi$ of a  WS channel. This is done by measuring the time-dependent intensity $I(t)$ transmitted by the WS channel and by computing the real part of the Fourier transform $\tilde{I}$ at frequency $\Omega_M-\Omega_D$ as a function of the heater power $P$. The data is fit with the sinusoidal relation $\sim\sin \left ( \frac{2\pi P}{P_{2\pi}} +\Phi_0\right )$, shown as a solid line. The 95$\%$ confidence bounds from the fit are indicated with shaded regions.}}
\label{fig:set-up}
\end{figure*}

The silicon photonic frequency processor, reproduced in Fig.\ref{fig:set-up}(b), is fabricated within a multi–project wafer run on the 300~mm Si300~nm photonics platform at STMicroelectronics \cite{2021_Boeuf_ST} (more details on the waveguide and component geometries are provided in the Methods section). The chip area occupied by the circuit is approximately $4\times7~\mathrm{mm}^2$ and integrates all the basic functional building blocks required to implement a QFP. The optical circuit is reconfigured by on-chip heater elements which are wire-bonded to an external printed circuit board (see Fig.~\ref{fig:set-up}(a)) and controlled by a driving electronics. The whole sample is thermally stabilized using a Peltier cell.\\ 
Following Fig.~\ref{fig:set-up}(b), a pump laser is fiber coupled to the IN-1 port of the chip by using grating couplers, where it excites a microresonator-based photon-pair source designed to emit a BPFC by spontaneous four-wave mixing with free spectral range (FSR) of $\simeq15.34$ GHz. The resonator has a racetrack geometry and is tuned into the critical coupling regime by using an interferometric coupler.
After the source, the pump is attenuated by means of an asymmetric Mach–Zehnder interferometer (MZI), which acts as a  pump rejection filter (PF) and thus avoiding spurious photon-pair generation in the rest of the circuit. The BPFC is then directed to the three element QFP, which akin to the architecture implemented in Ref.~\cite{2018_Lu_FB_BS} consists in line-by-line waveshaper (WS) sandwiched between an input and an output fast phase modulator (IN PM and OUT PM). 
The intuition behind the working principle of the QFP is sketched in Fig.~\ref{fig:set-up}(c). Each FB at the input is coherently mixed with its neighbors by the IN PM, then independent spectral phases are applied to each bin before a second mixing process occurs at the OUT PM. The phase profile applied by the WS, together with the relative phase and amplitude of the RF signals driving the PMs, determines how the frequency bins interfere. 
The process can be understood as a frequency-domain generalization of the MZI meshes based on beamsplitter networks and phase shifters used for reconfiguring the optical paths \cite{sharma2025universal}, with the important subtlety that the mode mixing operated by the PMs is intrinsically high-dimensional.\\ 
Specifically, mode mixing is implemented on chip by p-n traveling-wave carrier depletion modulators, each driven by a single tone RF signal whose frequency matches the FSR of the BPFC (see Methods section for fabrication and performance details).\\
The line-by-line WS consists of four cascaded pairs (WS-units) of add–drop microring resonators (MRR). The units can be operated in three different settings, as illustrated in Fig.~\ref{fig:set-up}(d). In the standard PHASE configuration, the unit targets a specific channel wavelength and is used to control the phase of the corresponding frequency bin. The DEMUX ring of each pair (DR in Fig.~\ref{fig:set-up}(d)) drops a specific wavelength, which is routed towards the add-port of the second resonator after experiencing a programmable phase-shift $\Phi$. The MUX ring (MR in Fig.~\ref{fig:set-up}(d)) adds back the dropped wavelength into the common bus waveguide, thus enabling independent phase control of the target comb line (Fig.~\ref{fig:set-up}(c)). To ensure a flat channel response across the BPFC modes, each ring has been designed to have a quality factor (Q) of about $5.2\times10^4$, which is approximately six times lower than the loaded Q of the source (see Methods section for further details). A comparison between the WS-unit transmission spectrum and a resonance of the source is reported in Fig.~\ref{fig:set-up}(e). In the PASS configuration, both rings are detuned from the channel wavelength and the FB is transmitted without any amplitude modulation. In the STOP configuration, either the first or the second ring resonance is aligned to the central channel wavelength, directing the light towards a Germanium-based absorber that blocks its propagation.\\
The calibration of each PHASE channel is done in two steps. First, the resonances of the two MRRs are aligned to the central channel wavelength. Then, the phase $\Phi$ is set to the target value. These steps are implemented using feedback signals derived from the amplitudes of higher-order harmonics in the transmitted intensity, which are generated by small dither tones applied to the voltages of the heaters of the two rings (see Fig.~\ref{fig:set-up}(f,g); a detailed description of the calibration procedure is provided in the Methods section). 

\subsection*{Frequency beamsplitters and single qubit gate synthesis}
\label{sec:Results_BS}
\begin{figure*}[!t]
\centering
\includegraphics[width = \textwidth]{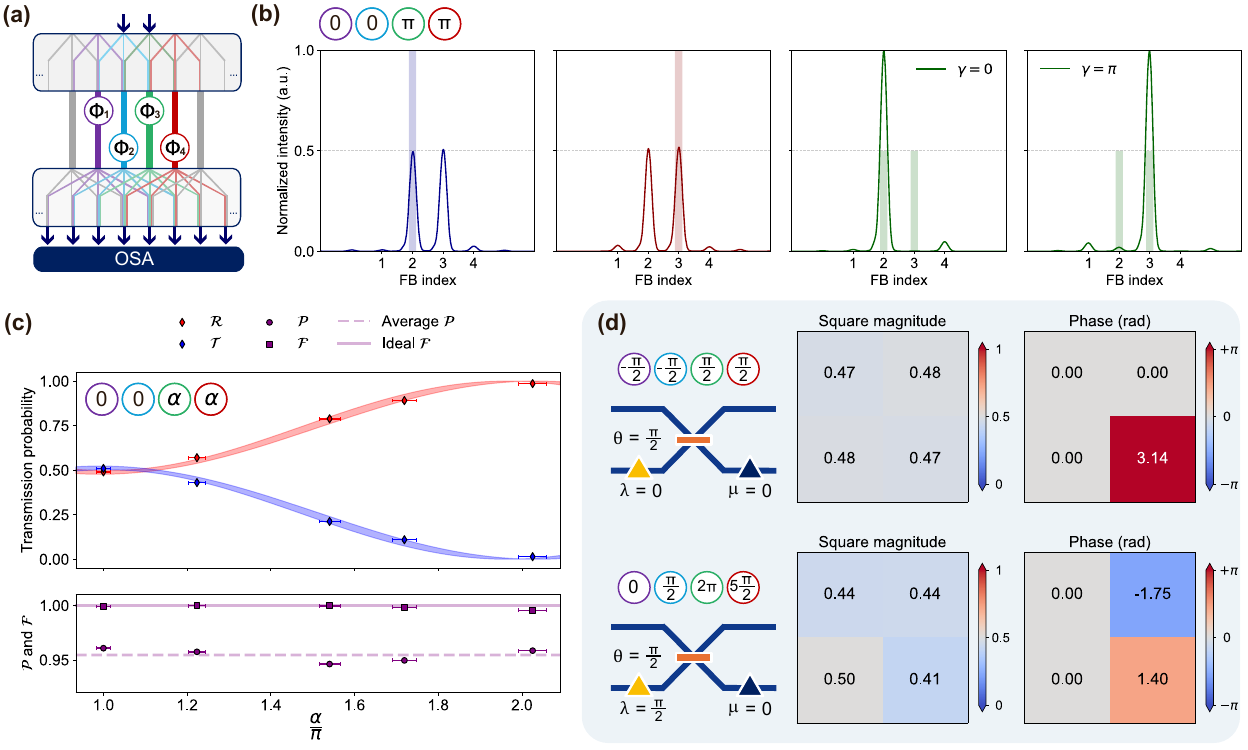}
\captionof{figure}{\parbox{\textwidth}{\noindent\justifying
\textbf{Implementation of frequency beamsplitters ($R_y$ rotations) with a tunable splitting ratios and $R_z$ rotations.} (a) Schematic of the processor configuration used to implement frequency beamsplitters with a tunable splitting ratio. An external probe laser is injected either into FB 2 or FB 3, or into a coherent superposition of the two. After the IN PM, programmable spectral phases $\{\Phi\}_{j=1}^4$ are applied to four modes before they enter the OUT PM. The spectral response at the output of the processor is recorded by an optical spectrum analyzer (OSA).
(b) Measured output spectra when the processor is programmed to realize a 50/50 beamsplitter. The four spectral phases $\{\Phi\}_{j=1}^4$ applied by the WS are shown in the top inset of the leftmost panel. From left to right: input (shaded vertical line) in FB 2, input in FB 3, input in a coherent superposition of FBs 2 and 3 with relative phase $\gamma=0$, and with a relative phase $\gamma=\pi$.
(c) Top: tunable beamsplitter reflectivity $\mathcal{R}$ and transmittivity $\mathcal{T}$ as a function of the relative phase $\alpha$ applied between the FBs 2 and 3. The shaded regions represent the 95$\%$ confidence bounds predicted by the best fit using Eq.(\ref{eq:RandT}). Bottom: success probability $\mathcal{P}$ of the transformation and fidelity $\mathcal{F}$ with respect to the target operation. Errors on $\alpha$ are derived from the power-to-phase calibration of $\Phi(P)$ shown in Fig.~\ref{fig:set-up}(g). The y-error bars are shown and are included within the marker size. 
(d) Top: reconstructed scattering matrix (squared magnitude and phase) of the 50/50 beamsplitter, corresponding to the operation $R_y(\theta=\pi/2)$. The equivalent transformation, illustrated using dual-rail encoding, is sketched to the left of the matrices. The spectral phases applied to the four FBs are indicated inside the colored circles (see correspondence with panel (a)). Bottom: reconstructed scattering matrix of the transformation $R_y(\theta=\pi/2)R_z(\lambda=\pi/2)$.}}
\label{fig:Results_BS}
\end{figure*}
We first use the processor to implement a fundamental building block for coherent frequency control: a deterministic beamsplitter between two frequency modes. This operation is nontrivial to implement in the frequency domain and cannot be achieved with unit success probability—defined as the fraction of light remaining confined to the two input modes after the beamsplitter operation and normalized by the device insertion loss—using phase modulators alone, as they induce unwanted scattering of light into FBs outside the two modes of interest.
However, as shown in Ref.~\cite{2018_Lu_FB_BS}, a three-element QFP, such as the one implemented on our chip and schematically depicted in Fig.~\ref{fig:Results_BS}(a), can perform this operation almost deterministically.\\ 
To realize this operation, we require the two PMs to be driven by RF waveforms with a relative phase shift of $\pi$. When no spectral phase is applied by the WS, the frequency bins are scrambled by the IN PM and recombined out of phase at the OUT PM, effectively yielding the identity operation. By contrast, introducing a relative phase $\alpha$ between the  frequency bins involved in the beasmplitter operation breaks the destructive interference at the OUT PM, enabling tunable splitting ratios. Crucially, destructive interference is simultaneously preserved for all non-targeted modes, thereby suppressing unwanted scattering (see Supplemental Document for mathematical details).\\
We characterize the beamsplitter operation using two target frequency bins separated by $\Delta\nu = 13.25$ GHz and centered at a wavelength of 1547~nm. The input probe light is prepared off-chip using an electro-optic frequency comb filtered by a programmable waveshaper that selects and equalizes the amplitudes of the input modes (see Methods section for the detailed off-chip setup). For this characterization, the photon-pair source is bypassed by injecting light through the IN-2 port of the chip (Fig.~\ref{fig:set-up}(b)). The IN and OUT PMs are driven by a single RF tone at 13.25~GHz with a relative $\pi$ phase shift and a modulation depth of $\delta \simeq 0.82\,$rad \cite{2018_Lu_FB_Q}.\\
Using the on-chip WS, we independently control the phases $\{\Phi_j\}_{j=1}^{4}$ of the four channels, labeled 1 through 4 in Fig.~\ref{fig:Results_BS}(a). The beamsplitter operation is implemented between the inner modes 2 and 3 by setting $\Phi_1 - \Phi_2 = \Phi_3 - \Phi_4 = 0$, and $\Phi_3 - \Phi_2 = \alpha$.\\
We initially focus on 50/50 operation, which is realized by setting $\alpha = \pi$ (see Supplemental Document and Ref.~\cite{2018_Lu_FB_Q}). In this condition, when light is separately injected into frequency bins 2 and 3, it is found equally split at the output, as shown in Fig.~\ref{fig:Results_BS}(b). To assess coherent operation, we prepare an equal superposition of the two modes at the input, with a relative phase $\gamma$ that is varied using the external phase modulator generating the electro-optic comb. Meanwhile, the output spectra is monitored with an optical spectrum analyzer. As shown in Fig.~\ref{fig:Results_BS}(b), the output light can be fully routed to bin 2 or bin 3 by setting $\gamma = 0$ and $\gamma = \pi$, respectively, demonstrating the coherent mode mixing of the frequency beamsplitter.\\
We then vary $\alpha$ from $\pi$ to $2\pi$ to tune its reflectivity $\mathcal{R}(\alpha)$ and transmissivity $\mathcal{T}(\alpha)$. Theoretically, $\mathcal{R}$ can be tuned from a minimum of 0.4978 by setting $\alpha = \pi$ to a maximum of 1 when $\alpha = 2\pi$, while $\mathcal{T}$ varies from a maximum of 0.4781 for $\alpha = \pi$ to zero when $\alpha = 2\pi$, following a cosinusoidal dependence (see Supplemental Document for details). The measured $\mathcal{R}(\alpha)$ and $\mathcal{T}(\alpha)$, shown in Fig.~\ref{fig:Results_BS}(c), are in good agreement with theory. Crucially, the success probability remains above $94\%$ over the entire range $\alpha \in [\pi, 2\pi]$, well above the $\sim66\%$ upper bound achievable with a single phase modulator.\\
The high success probability is further evidenced by the spectra shown in Fig.~\ref{fig:Results_BS}(b), where the light intensity scattered into modes outside the target bins (2 and 3) is consistently at least an order of magnitude lower than that within the target modes.\\
We use the four spectra shown in Fig.~\ref{fig:Results_BS}(b) to reconstruct the complex $2\times2$ scattering matrix of the 50/50 beamsplitter \cite{2018_Lu_FB_BS}, shown in Fig.\ref{fig:Results_BS}(d). The fidelity, defined as the Frobenius overlap with the ideal target transformation, is $\mathcal{F} = 0.999(1)$.\\
Having validated the beamsplitter transformation for different splitting ratios, which within a quantum mechanical picture can be described by the parametrized rotation $R_y(\theta)=e^{i\theta Y}$ (with $Y$ denoting the corresponding Pauli operator) on the Bloch sphere, we then synthesize more general single-qubit gates. An arbitrary unitary operation on two modes would require two additional rotations, $R_z(\lambda)=e^{i\lambda Z}$ and $R_z(\mu)=e^{i\mu Z}$ (with $Z$ denoting the corresponding Pauli operator), applied before and after $R_y(\theta)$. Without loss of generality, we focus on the specific setting $\theta=\pi/2$ (50/50 splitting ratio), $\lambda=\pi/2$, and $\mu=0$, but arbitrary rotation angles can be similarly implemented. \\
As shown in Ref.~\cite{2020_Lu_fully}, the operation $R_y(\theta)R_z(\lambda)$ can be realized by starting from the processor settings that implement $R_y(\theta)$ and then adding a phase shift $-\lambda$ to the RF signal driving the IN PM, together with a linear spectral phase ramp applied to the WS,  i.e. $\Phi_k \mapsto \Phi_k + k\lambda$ (similarly, the final rotation $R_z(\mu)$ could be implemented by adding a phase shift $-\mu$ to the OUT PM and by updating the linear phase ramp as $\Phi_k \mapsto \Phi_k + k(\lambda+\mu)$). We applied this procedure and reconstructed the complex scattering matrix, which is shown in Fig.~\ref{fig:Results_BS}(d) (bottom panel). The fidelity of this matrix with respect to the target is $0.993(1)$, and the operation is performed with a success probability of $0.90(1) $.\\
These results demonstrate that, through accurate phase control over four comb lines, the device performs reproducible and high-fidelity single-qubit gates without significant population of frequency bins outside the computational subspace. Although the success probability could be marginally improved by controlling additional side channels with the WS, such gains would come at the cost of increased system complexity  \cite{2018_Lu_FB_BS}. 
As a final remark, while the $R_y(\theta)$ rotation shown in Fig.~\ref{fig:Results_BS}(c) is demonstrated over the range $\theta \in [0,\pi/2]$, corresponding to $\alpha \in [\pi,2\pi]$, it could in principle be extended to the full range $\theta \in [0,\pi]$ by further increasing the modulation depth $\delta$. However, as discussed in Ref.~\cite{2020_Lu_fully}, larger values of $\delta$ necessarily reduce the success probability unless additional WS channels and RF tones are employed, which are beyond the capabilities of the processor in its current implementation.\\

\subsection*{Tunable quantum walks and quantum state tomography}
\label{sec:Results_Q}
\begin{figure*}[!t]
\centering
\includegraphics[width = \textwidth]{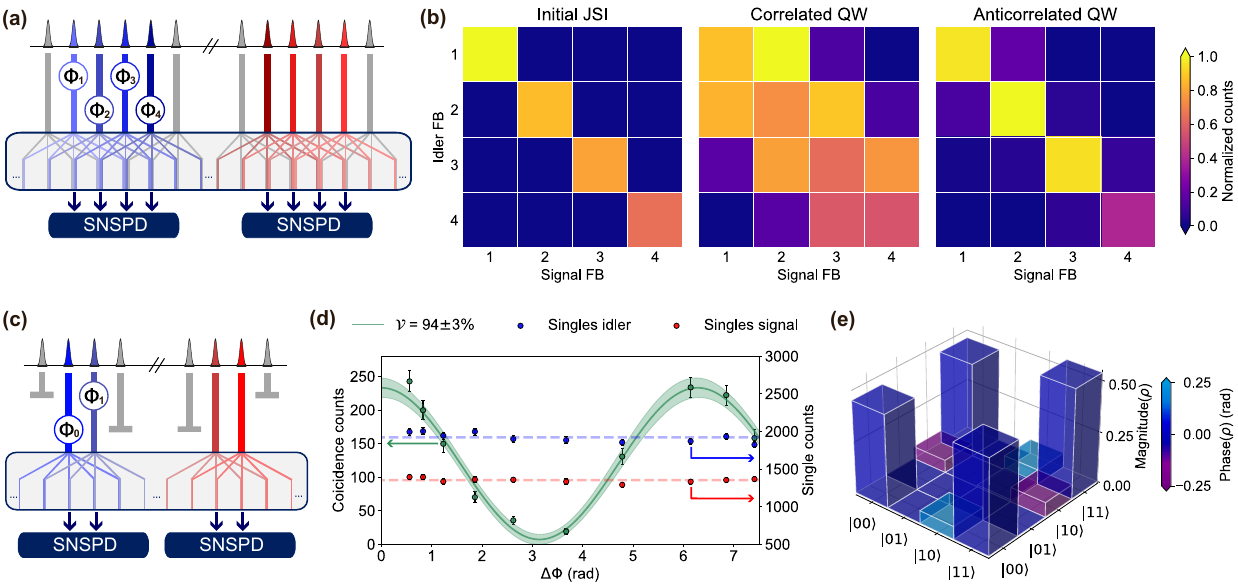}
\captionof{figure}{\parbox{\textwidth}{\noindent\justifying
\textbf{On-chip coherent manipulation of high-dimensional frequency-bin entangled states and quantum state tomography.} (a) Schematic of the processor configuration used to implement high-dimensional quantum walks. A programmable spectral phase is applied to four idler (blue) modes, labeled from 1 to 4. Signal and idler modes are independently mixed at the OUT PM. Frequency-resolved signal–idler coincidence events are recorded off chip using superconducting nanowire single photon detectors (SNSPD).
(b) JSIs in a restricted $4\times4$ subspace in three different processor configurations. From left to right: the initial JSI of the source, after the correlated quantum walk, and after the anticorrelated quantum walk. The recorded coincidence counts are here reported normalized to the maximum of each subspace. 
(c) Processor configuration used to implement on-chip two-qubit quantum state analysis. Two signal and two idler comb lines are isolated by guard-band modes. Spectral phases are applied to the idler FBs to control the measurement projectors, while the OUT PM enables projections onto superposition bases.
(d) On-chip Bell state analysis. The measured coincidences (green) and single counts (blue and red) are reported as a function of the relative phase $\Delta\Phi = \Phi_1-\Phi_0$ introduced between the FBs by two WS channels. The green curve (solid line) is a sinusoidal fit of the data, where 95$\%$ confidence bounds are shown as shaded regions.
(e) Reconstructed magnitude (height of the bars) and phase (color of the bars) of the density matrix through quantum state tomography of the two-qubit state, which closely approximates the maximally entangled Bell state $\ket{\Phi^{+}}$.}}
\label{fig:Results_Q}
\end{figure*}

We now demonstrate the capability of the integrated QFP to manipulate the BPFC generated by the on-chip source. To this end, the microresonator is resonantly excited with a pump power of 2.5~mW, generating signal-idler photon pairs with an on-chip rate of $\sim 1.0$ MHz and a coincidence-to-accidental ratio (CAR) of $55 \pm 12$ for each resonance pair. In what follows, we will restrict the frequency-resolved coincidence measurements to the $4 \times 4$ subspace spanned by the four idler frequency bins passing the WS channels (labeled 1-4 in Fig.\ref{fig:Results_Q}(a)) and their energy-correlated signal bins, due to the limited bandwidth ($\sim 80$ GHz) of the filters implemented off-chip for idler-signal separation and pump suppression.
Figure~\ref{fig:Results_Q}(b) (Initial JSI panel) shows the frequency-resolved coincidence measurement over this sixteen dimensional space, i.e the joint spectral intensity (JSI) immediately after the source. No background noise from photon pairs generated off-resonance is observed, confirming efficient suppression by the pump-rejection filter.\\ 
To demonstrate coherent manipulation of this high-dimensional state, we apply a single RF tone at $15.34$~GHz, matching the FSR of resonator, with a modulation depth of $\delta \simeq 0.82\,$rad to the OUT PM, while keeping the IN PM off and the four WS channels in the PASS configuration. The applied RF signal coherently mixes FBs within both the signal and idler subspaces, effectively inducing a frequency-domain quantum walk whose depth increases with $\delta$.\\
As discussed in Ref.~\cite{2020_Imany_QWalk}, the efficiency of energy exchange between the modulator and the photon pair depends critically on the relative phase $\Delta = \textrm{Arg}(\beta_{l+1}) - \textrm{Arg}(\beta_{l})$ between adjacent frequency bins prior to the walk: is maximized for $\Delta = 0$ and minimized for $\Delta = \pi$, corresponding to correlated (ballistic) and anticorrelated (strongly confined) quantum walks, respectively. The biphotons generated by the MRR naturally exhibit $\Delta \approx 0$ (see numerical simulations in the Supplemental document), and therefore by turning on the OUT PM they undergo a correlated walk, with frequency correlations spreading away from the main diagonal as shown in Fig.~\ref{fig:Results_Q}(b) (correlated QW  panel). We then convert the walk from correlated to anticorrelated by applying a relative phase $\Delta = \pi$ between adjacent idler frequency bins using the integrated WS. The resulting JSI, shown in Fig.~\ref{fig:Results_Q}(b) (Anticorrelated QW panel), clearly demonstrates an increased confinement of the biphoton energy along the main diagonal. The WS is essential for this operation because both photons propagate through the same PM. Routing the idler and signal photons through separate PMs would allow an equivalent phase shift to be introduced by delaying one of the RF drive signals; however, this approach would require additional modulators, RF lines, and spectral filtering stages, thereby increasing circuit complexity and footprint.\\
Finally, we exploit the independent control of signal and idler photons to perform on-chip quantum state tomography of two frequency-encoded qubits. The state is carved from the broadband BPFC by creating guard bands around two signal and two idler frequency bins, labeled $0$ and $1$ in Fig.~\ref{fig:Results_Q}(c), yielding a state well approximated by the Bell state $\ket{\Phi^{+}}=\frac{1}{\sqrt{2}}\left ( \ket{00}_{si}+\ket{11}_{si} \right)$. Guard bands are implemented using four MRRs of the WS (STOP configuration), providing approximately 12$\sim$15~dB suppression of the unwanted modes. The remaining four MRRs, forming two WS channels in the PHASE configuration, apply programmable spectral phases to the idler modes while leaving the signal modes unaffected (PASS configuration). With the OUT PM turned on and by performing a frequency resolved coincidence measurement on $\ket{00}_{si}$ as a function of the relative phase $\Delta\Phi =\Phi_1-\Phi_0$ between the idler FBs,  we observe the two-photon interference fringe reported in Fig.~\ref{fig:Results_Q}(d). The visibility obtained from the fit is $94(3)\%$, violating the Bell inequality by more than 7 standard deviations.\\ 
Furthermore, a set of projective measurements is realized to perform on-chip quantum state tomography by combining spectral phase control with frequency mixing via the OUT PM (the latter turned off for projections on the computational basis). The density matrix, reconstructed through maximum-likelihood \cite{2023_Borghi}, is shown in Fig.~\ref{fig:Results_Q}(e). The measured fidelity with respect to the target Bell state $\ket{\Phi^+}$ is $0.957(3)$, and the purity $0.938(5)$.

\section*{Discussion}
We demonstrated key elements required for frequency-bin quantum information processing, namely quantum state generation, coherent frequency mixing, and spectral phase control. This is enabled by the monolithic integration and the high device uniformity provided by the silicon photonic platform. We observed highly reproducible performance across all processor configurations, calibrated through electronic and thermal control parameters, demonstrating that complex linear transformations between frequency bins can be implemented all on-chip with high accuracy and precision. 
In addition, the strong third-order nonlinearity of silicon and tight optical mode confinement enable the co-integration of an on-chip photon-pair source, while fast carrier-depletion modulators provided the required frequency-mixing capability. The waveshaper architecture proposed here, in contrast to recent integrated microring-based pulse shapers \cite{cohen2024silicon}, is intrinsically non-blocking, preserving all frequency bins on a single bus waveguide and enabling lossless pass-through operation for selected modes while coherently manipulating others. In perspective, the QFP architecture could address a larger number of modes by adding MRR pairs to the line-by-line WS, thereby enabling the implementation of parallelizable gates as demonstrated with bulk components in Ref.~\cite{henry2024parallelization}.\\
Universal transformations on $N$ modes would require $\mathcal{O}(N)$ WS-PM blocks (more precisely, in Ref.~\cite{2016_Lukens_QFP_theory} it is shown that an upper bound on the number of components is $2N-1$, while determining the minimum number of elements remains an open problem and currently relies on numerical optimization). Therefore, integration is a necessary step toward the realization of large-scale QFP, and our work marks important progress in this direction. Equally important is the availability of a clear strategy to reduce the loss associated with each component. At present, the waveshaper contributes $\sim6\,$dB of loss and currently represents the main performance limitation of the architecture, although this value is consistent with similar integrated implementations \cite{cohen2024silicon}. However, this can be substantially mitigated by increasing the coupling coefficient between the MRRs and the bus waveguides, or by employing asymmetric couplers to improve the drop efficiency. Phase modulators based on carrier depletion suffer from intrinsic carrier absorption, which typically introduces an insertion loss of $0.5$–$1$ dB for a $1$-mm-long device \cite{rahim2021taking}. Alternatives are actively being explored, including the integration of CMOS-compatible electro-optic polymers \cite{taghavi2022polymer} and the heterogeneous integration of thin-film lithium niobate on silicon \cite{han2025heterogeneously}. It is worth noting that other architectures have been proposed that may reduce the loss and the resource cost of the QFP scheme. For example, a theoretically proposed approach employs $N$ cascaded resonators, each driven by $N$-RF tones spaced by the resonator FSR, to implement arbitrary unitary transformations on $N$ modes \cite{buddhiraju2021arbitrary}. This strategy eliminates the need for a waveshaper, shifting the system complexity to RF waveform generation and mixing.
Another promising approach exploits coupled resonators for frequency-domain beamsplitter operations \cite{hu2021chip}. Such devices can be arranged into a universal triangular architecture with only nearest-neighbor mode coupling \cite{reck1994experimental}, an idea that has been partially explored in variational circuits for multimode squeezed-light decomposition \cite{karnieli2025variational}. This strategy is particularly appealing because it can be implemented using compact microring resonators with large FSRs and requires only a single RF tone matched to the frequency separation of the supermodes of the photonic molecule \cite{hu2021chip}.\\
Near-term silicon-based frequency processors may therefore be well suited for protocols and applications that do not require ultra-low-loss operation. A notable example is decoy-state quantum key distribution (QKD), where silicon photonic transceivers based on time-bin or polarization encoding have already demonstrated excellent performance \cite{bunandar2018metropolitan}, as well as related protocols such as coherent one-way QKD \cite{sibson2017integrated} and measurement-device-independent QKD \cite{wei2020high}.
Equally compelling are frequency processors operating outside the single-photon regime, with applications including pulse shaping, neuromorphic computing, and complex-field spectrum analysis. 
\\

\subsection*{Funding and acknowledgments}
S.C. and D.B. acknowledge European Union funding from the STARLight project (project ID: 101194170). M.B.
acknowledges the PNRR MUR project PE0000023-NQSTI. M.G. acknowledges European Union funding from the HyperSpace project (project ID 101070168).\\
The authors declare no competing interests.\\

\section*{Materials and methods}
\subsection*{Detailed description of the processor layout}
The silicon photonic chip is fabricated within a multi-project wafer using the PIC50 technology of STMicroelectronics. The waveguides are implemented in single-mode RIB geometry with a 300~nm-thick silicon core, a 150~nm partial etch, and widths of 450~nm (core) and 3050~nm (slab). The propagation losses are estimated to be $\sim1.2$ dB/cm. Light is coupled into and out of the chip through grating couplers, which have an insertion loss of $3.0\pm0.5$ dB at $1550$ nm.\\
The source consists of a $\sim$5.1~mm-long resonator whose coupling coefficient to the bus waveguide can be tuned via thermal phase shifters. This is achieved using a symmetric MZI in a push–pull configuration. The experiments are performed close to the critical-coupling regime, corresponding to an average quality factor of $3\times10^5$ over the resonances of interest and a measured brightness per resonance of $298 \pm 1~\mathrm{kHz/mW^2}$. The source is followed by a pump rejection filter (PF in Fig.~\ref{fig:set-up}(b)) based on an asymmetric MZI with a free spectral range of 500~GHz. The filter provides approximately 30~dB of attenuation at the pump wavelength (1555.9~nm), while the signal and idler resonances exploited in the demonstration of Fig.~\ref{fig:Results_Q} are located about 5~nm away from the pump, at approximately 1550.3~nm and 1561.6~nm, respectively. The filter offers additional input and output ports for calibration and testing (IN-2 and OUT-2 in Fig.~\ref{fig:set-up}(b)).\\
Four WS-units are included in the circuit, whose design and operating principle is shown in Fig.~\ref{fig:set-up}(d). The central wavelength of each WS-unit and the phases $\{\Phi\}_{l=1}^4$ are controlled through thermal phase shifters. In the PHASE configuration, the non-unit drop efficiency results in an insertion loss of $5\sim6\,$dB per channel, with a small dependence on the applied phase $\Phi$ (see the experimental spectra in the Supplemental document). The loss values are consistent with similar integrated architectures \cite{cohen2024silicon} and improvable by increasing the coupling coefficient of the MRRs.\\
The chip integrates two silicon carrier-depletion modulators, referred to as IN PM and OUT PM in Fig.~\ref{fig:set-up}(b), based on 3~mm-long p–n junctions embedded in a deep-RIB waveguide geometry with a slab thickness of 50~nm and a core width of 360~nm, in which the doping regions are specifically engineered for high-speed operation. More details on the cross section and fabrication process can be found in Ref.~\cite{2019_Boeuf_ST,2021_Boeuf_ST}. The two modulators are integrated side by side, with light co-propagating in the same direction, enabling RF excitation using a single two-signal RF probe. A common traveling-wave electrode, designed in a signal–ground–signal (SGS) configuration, is integrated above the two junctions and features a line impedance of $50~\Omega$ with load termination. Each PM is driven by an independent RF signal and is simultaneously biased at $-2$~V through a bias-tee to ensure reverse-bias operation. The estimated insertion loss of each modulator is approximately $\sim1.5$~dB at $1550\,$nm. The measured junction modulation efficiency is $6.45\pm0.15~^\circ\mathrm{mm}^{-1}\mathrm{V}^{-1}$ in the DC regime; strategies to increase the modulation efficiency in the present photonic platform can be found in Ref.~\cite{2020_Boeuf_STopt}.\\
In the experimental demonstrations here reported, the modulators are driven by single-tone RF signals with electrical powers $\simeq  27$--$30$~dBm, corresponding to modulation depths in the range $\delta=0.8$--$1.4\,$rad. These power levels are consistent with the measured electrical loss of the traveling-wave electrode at the operating frequencies (13--15~GHz), which is approximately $9\,$dB. 
The relative RF phase between the two PMs is calibrated by setting each WS-unit to the PASS configuration and injecting a continuous-wave laser at the processor IN-2, with both the input and output PMs driven at equal modulation depth. The RF phase of the output PM is then varied until the electro-optic comb generated by the input PM is mapped back onto the carrier frequency by the second PM. This condition occurs for a relative RF phase of $\pi$, which is then used for phase reference.\\

\subsection*{Detailed description of the setup}
Here we provide a detailed description of the off-chip setup, whose schematic is included in the Supplemental Material for brevity. A tunable CW pump laser is sent to an off-chip electro–optic phase modulator driven by an RF signal at the modulation frequency used in the experiment (13.25 or 15.34~GHz). The resulting electro–optic comb lines are then selected by configuring an off-chip waveshaper as a passband filter centered on the target input-bin frequency. After polarization optimization and, when required, suppression of residual broadband spectral components using a bandpass filter (BPF), the light is injected into either the IN-1 or IN-2 port shown in Fig.~\ref{fig:set-up}. The on-chip operations are reconfigured by tuning the DC bias applied to the thermal phase shifters using dedicated computer-controlled voltage drivers. During the calibration of a single WS unit, dither tones are applied to its MRRs to track their resonance positions and induced phase shifts (see the next section for details of the calibration procedure).\\
By adjusting the thermal phase shifter of the PF, different on-chip wavelengths can be routed either toward the rest of the chip or directly coupled off chip through the OUT-2 port. For gate synthesis and calibration of the WS channels, the pump laser is injected into IN-2, thereby bypassing the source, and routed toward the QFP. After frequency-bin manipulation by the PMs and the WS, the light is coupled off chip and measured either with an OSA, for gate characterization, or with a photodiode, for WS channel calibration.\\
For the quantum walk and quantum state tomography demonstrations, the pump light is instead injected into IN-1, tuned into resonance with the source, and routed off chip through OUT-2. The transmitted laser light is monitored with a photodiode, and the pump laser current is adjusted in an active feedback loop to maintain the laser wavelength locked to the center of the pump resonance. The generated BPFC, shaped by the sine-like transmission spectrum of the PF, is directed toward the QFP and subsequently manipulated by the WS and the OUT PM. Signal and idler photons are then collected from the OUT-1 port and directed to a frequency-resolved single photon detection setup, based on DWDM (bandwidth $\sim80\,$GHz) and tunable narrowband (bandwidth $\sim12\,$GHz) filters.

\subsection*{Calibration of the line-by-line waveshaper}
\label{sec:WS_Cal}
The calibration of the WS follows two main steps.\\
\emph{Alignment of the ring resonances}\\
In the first step, the resonances of the DR and MR rings of in Fig.~\ref{fig:set-up}(d) are aligned to the central channel wavelength $\lambda_{\mathrm{ch}}$. This is achieved by applying two low-amplitude dither tones at frequencies $\Omega_D = 150 \,\mathrm{Hz}$ and $\Omega_M = 250\,\mathrm{Hz}$ to the heaters of the DR and MR rings, respectively, thereby inducing a small modulation of their resonance wavelengths with amplitude $\Lambda$. The resulting time-dependent transmitted intensity $I(t)$ is recorded using a photodiode, and its spectral content is extracted via a fast Fourier transform (FFT).\\
The strongly nonlinear spectral response of the WS channel around the probe wavelength generates multiple harmonics in the FFT spectrum. The complex amplitudes of these harmonics depend on the relative detunings $\Delta \lambda_{D(M)} = \bar{\lambda}_{D(M)} - \lambda_{\mathrm{ch}}$ between the average resonance wavelengths of the DR(MR) rings, $\bar{\lambda}_{D(M)}$, and the channel wavelength $\lambda_{\mathrm{ch}}$, as well as on the phase $\Phi$ imparted by the WS channel.\\
Using numerical simulations based on standard scattering-matrix formalism (see Supplemental document for details) —in which the resonance wavelengths of the DR and MR rings are modulated as $\lambda_{D(M)}(t) = \bar{\lambda}_{D(M)} + \Lambda \sin(\Omega_{D(M)} t)$—we find that the magnitude of the harmonic at frequency $2(\Omega_D + \Omega_M)$ is maximized when $\Delta\lambda_M = \Delta\lambda_D = 0$, independent of the value of $\Phi$. This condition therefore provides a robust criterion for aligning the resonances of each ring pair within a WS channel.\\
In practice, the alignment procedure begins by injecting a laser at the target $\lambda_{\mathrm{ch}}$ by the IN-2 port, and routing the light to the WS. After a coarse tuning of both ring resonances near $\lambda_{\mathrm{ch}}$, a grid search in heater currents is implemented to maximize the amplitude of the $2(\Omega_D + \Omega_M)$ harmonic. An example of such a scan is shown in Fig.~\ref{fig:set-up}(f). At the conclusion of the scan, the heater currents are set to the values that maximize this harmonic amplitude, thereby achieving optimal and automatic resonance alignment.\\  
\emph{Mapping of the heater power to the phase $\Phi$}\\
The second step consists of determining the relationship between the electrical power $P$ applied to the heater located between the DR and MR of a WS-unit and the induced spectral phase $\Phi$. Numerical simulations show that, when the resonances are aligned ($\Delta\lambda_{D(M)} = 0$), the real part of the harmonic component at frequency $\Omega_M - \Omega_D$ follows the simple relation $\textrm{Re}(\tilde{I}(\Omega_M-\Omega_D))=\tilde{I}_0\cos(\Phi+\Phi_0)$, where $\Phi_0$ is a constant phase offset corresponding to zero applied electrical power.
Accordingly, after aligning the ring resonances as described in the previous step, we sweep the electrical power $P$ applied to the heater and record a full  fringe in $\mathrm{Re}\left[\tilde{I}(\Omega_M - \Omega_D)\right]$. The measured fringe is then fitted using $\textrm{Re}\left [\tilde{I}(\Omega_M-\Omega_D) \right ]=\tilde{I}_0\cos(2\pi P/P_{2\pi}+\Phi_0)$, where $P_{2\pi}$ and $\Phi_0$ are treated as free fitting parameters. An example of such a fringe is shown in Fig.~\ref{fig:set-up}(g).
This procedure completes the phase calibration, as it establishes the mapping between the applied electrical power $P$ and the induced phase, i.e  $\Phi(P)=\Phi_0+2\pi P/P_{2\pi}$.\\  
Both calibration steps can be parallelized across multiple wavelength channels. To this end, an electro-optic frequency comb with the desired channel spacing can be generated either off chip or directly on chip using the IN PM, and subsequently demultiplexed at the WS output to provide a distinct signal for each WS-unit. In this configuration, the same pair of dither tones can be applied to the rings associated with each channel without introducing crosstalk.\\
Alternatively, distinct pairs of dither tones can be assigned to different WS channels, thereby eliminating the need to demultiplex the output signal into separate wavelengths. In this case, however, the dither frequencies must be carefully chosen to ensure that different WS-units do not produce identical harmonics.\\
For the proof-of-principle demonstration reported here, we did not exploit multiplexed calibration. Instead, we dithered one WS-unit at a time, sequentially calibrating the four WS channels and iterating the procedure to improve calibration precision.

\section*{Supplemental document}
\subsection*{Implementing single qubit rotations}
The most general unitary operation acting on a single qubit is described by the matrix
\begin{equation}
    U(\theta,\lambda,\mu)=
    \begin{pmatrix}
    \cos \left(\frac{\theta}{2}\right) & e^{i\lambda}\sin\left(\frac{\theta}{2}\right) \\
    e^{i\mu}\sin\left(\frac{\theta}{2}\right) & -e^{i(\lambda+\mu)}\cos\left(\frac{\theta}{2}\right)
    \end{pmatrix},
	\,    
\end{equation}
and can be decomposed into three sequential rotations $R_z(\lambda)R_y(\theta)R_z(\mu)$, parametrized by the angles $(\theta,\lambda,\mu)$ where $R_J(x)=e^{ixJ}$ and $J=\{Y,Z\}$ are the Pauli matrices.
We initially focus on the case $\mu=\lambda=0$. The input space of frequency bins is infinite-dimensional, but we restrict our attention to the $2\times2$ subspace describing the coherent mixing between two modes of interest, that we label $0$ and $1$ and are spaced by a frequency separation $\Delta\nu$. In its simplest implementation, a three element QFP has been shown to realize arbitrary single qubit rotations by properly engineering the modulation depths of the two modulators, the relative phase of the RF single tones, and the spectral phases imposed by the WS  \cite{2020_Lu_fully}. Specifically, if the relative phase between the PMs is set to $\pi$ and their modulation depth is set equal to $\delta$, the splitting ratio of the beasmplitter can be controlled by a single parameter $\alpha$, which is the relative phase between the bins $0$ and $1$ applied by the WS. In this configuration, all the bins $j<0$ and $j>1$ outside the computational basis have a zero relative phase offset, and the $2\times2$ matrix $V$ acting on the subspace of bins $0$ and $1$ writes \cite{2018_Lu_FB_Q}
\begin{equation}
    V(\alpha)= 
    \begin{pmatrix}
        \sqrt{\mathcal{R}(\alpha)} & \sqrt{\mathcal{T}(\alpha)} \\
        \sqrt{\mathcal{T}(\alpha)} & -\sqrt{\mathcal{R}(\alpha)}
    \end{pmatrix},
    \label{eq:Vmatrix}
\end{equation}
where the reflectivity $\mathcal{R}(\alpha)$ and transmittivity $\mathcal{T}(\alpha)$ are
\begin{align}
\label{eq:RandT}
\mathcal{T} &=
\left| \left(1 - e^{i\alpha}\right)\sum_{k=1}^{\infty} 
J_k(\delta) J_{k-1}(\delta) \right|^2 , \notag \\[6pt]
\mathcal{R} &=
\left| J_0^2(\delta) 
+ (1 + e^{i\alpha})\,\frac{1 - J_0^2(\delta)}{2} \right|^2,
\end{align}
and $J_k(\delta)$ is the Bessel functions of the first kind of order $k$. Note that the matrix in Eq.(\ref{eq:Vmatrix}) is in general not unitary, reflecting the fact that some light can be scattered outside the computational space. These expressions can be simplified by simple algebraic steps, yielding
\begin{align}
\label{eq:R_T_vs_alpha}
\mathcal{R}(\alpha) = & J_0^4+\left ( \frac{1-J_0^4}{2} \right)(1+\cos \alpha) \notag \\
\mathcal{T}(\alpha) = & \bar{J}(1-\cos \alpha)
\end{align}
where $\bar{J}=2|\sum_{k=1}^{\infty}J_k(\delta)J_{k-1}(\delta)|$. The 50/50 configuration is realized by setting $\delta=0.8169$, for which $\bar{J}=0.239$ and $V\approx U(\theta=\frac{\pi}{2},0,0)=R_y(\frac{\pi}{2})$ \cite{2018_Lu_FB_Q}.\\
The construction of arbitrary gates requires two additional $Z$ rotations, one before $R_y(\theta)$, parametrized by the angle $\lambda$, and the other after the Y rotation, parametrized by the angle $\mu$. In Ref.~\cite{2018_Lu_FB_Q} the authors found a simple rule that relates $U(\theta,0,0)$ to $U(\theta,\lambda,\mu)$: delay the RF signal of the first and second PM by $\tau_a=-\frac{\lambda}{\Delta\nu}$ and $\tau_b=-\frac{\mu}{\Delta\nu}$ respectively, and add the linear spectral phase ramp $\phi_k\mapsto \phi_k+k(\lambda+\mu)$ to each WS channel $k$. This is the procedure that we implemented in our work (see Fig.2(d) of the main text) to realize the operation $R_z(\frac{\pi}{2})R_y(\frac{\theta}{2})$ from an initial configuration that realizes $R_y(\frac{\theta}{2})$, i.e a 50/50 beamsplitter.\\ 
Two figures of merit are employed to verify the quality of the implemented transformations: the fidelity $\mathcal{F}$, which evaluates the overlap between the (experimentally reconstructed) matrix $V$ with the target $U_t$, and the success probability $\mathcal{P}$, which quantifies the fraction of light intensity remaining in the two target modes with respect to the total spectral intensity. Following Ref.~\cite{2018_Lu_FB_BS}, they are calculated using the Frobenius norm as:
\begin{align}
\mathcal{P} &=\frac{1}{2}\,\mathrm{Tr}\!\left(V^\dagger V\right) , \\[6pt]
\mathcal{F} &=\frac{1}{4\mathcal{P} \,} \left|\mathrm{Tr}\!\left(V^\dagger U_{t}\right)\right|^2 \, ,
\label{ch:05_eq:FoMs}
\end{align}
Note that since we are interested in evaluating the suppression of light into the unwanted modes $\neq (0,1)$, the success probability $\mathcal{P}$ is normalized by the insertion loss of the device in the main text. 

\subsection*{Additional figures}
\begin{widetext}
\begin{center}
\includegraphics[width=1.0\textwidth]{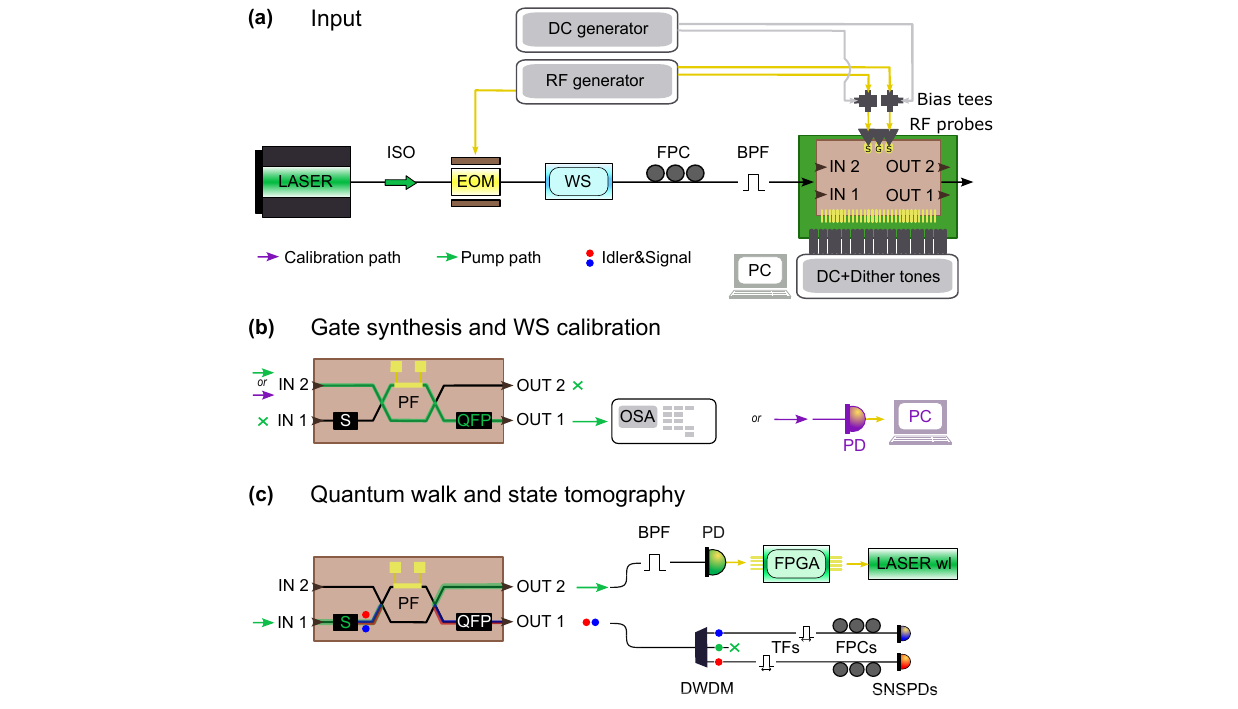}  
\captionof{figure}{\textbf{Detailed experimental setup.} (a) Setup at the input of the silicon chip, showing the main optical and electrical (DC bias, dither tones and RF signal) connections. (b) Configuration of the chip and setup used for gate synthesis and WS calibration (PHASE operation). (c) Configuration of the chip and setup used for quantum walks and quantum state tomography demonstrations.
ISO: fiber isolator, EOM: electro-optic modulator, WS: bulk off-chip waveshaper, FPC: fiber polarization controller, BPF: band pass filter, S: Source, PF: pump filter, QFP: quantum frequency processor, OSA: optical spectra analyzer, PD: photodiode, DWDM: dense wavelength division multiplexing, TF: tunable narrowband pass-band filter, SNSPD: superconducting nanowire single photon detector.
}
\label{fig:WS_cal}
\end{center}
\end{widetext}

\subsection*{Waveshaper characterization and simulation}
In Fig. \ref{fig:WS_cal}(a) we report the measured spectra around the central channel wavelength (here corresponding to zero-wavelength shift, as reported in the figure x-axis) of one WS-unit for different applied phases $\Phi$. Outside the target wavelength, part of the light which is transmitted from the through port of the DR to the input of the MR interferes with the light that is loaded back by the MR from the its add-port. The interference distorts the transmitted intensity, which becomes a function of $\Phi$, a feature that is exploited in the calibration phase discussed in the main text.\\ 
We simulated the time-dependent transmitted intensity of the unit while applying two independent dither tones at frequencies $\Omega_{D}$ and $\Omega_{M}$ to the two resonators, respectively DR and MR, as performed during the WS calibration procedure described in the Methods section of the main text. The simulations consider a pair of identical symmetric add–drop MRRs arranged in the here introduced WS-unit geometry, with a radius of approximately $50\,\mu$m, an extracted power coupling coefficient of 0.023, a waveguide propagation loss of $1.2\,$dB cm$^{-1}$, and a simulated effective index of the waveguide of $n_{\mathrm{eff}}\simeq2.8$.\\
The real and imaginary parts of the Fourier components of the time-dependent optical signal are recorded for different values of the MUX resonance detuning $\Delta\lambda_M$ and DEMUX resonance detuning $\Delta\lambda_D$. Figure~\ref{fig:WS_cal}(c) reports the results obtained for the harmonic at frequency $2(\Omega_D + \Omega_M)$ for four different channel phase shifts. While the overall magnitude pattern depends on the applied phase, its absolute maximum is located precisely at $\Delta\lambda_M = \Delta\lambda_D = 0$. In particular, the right panel corresponding to a phase shift of $\pi$ reproduces the experimental two-dimensional map shown in the main text.\\
Figure~\ref{fig:WS_cal}(b) shows the simulated real part of the harmonic at frequency $\Omega_M - \Omega_D$ as a function of the applied phase, calculated in the absence of detuning. The harmonic amplitude exhibits the expected $\cos(\Phi)$ dependence, which is used to map the heater thermal power to the corresponding induced phase shift, as described in detail in the Methods section. The corresponding experimental results are reported in the main text.
\begin{widetext}
\begin{center}
\includegraphics[width=\textwidth]{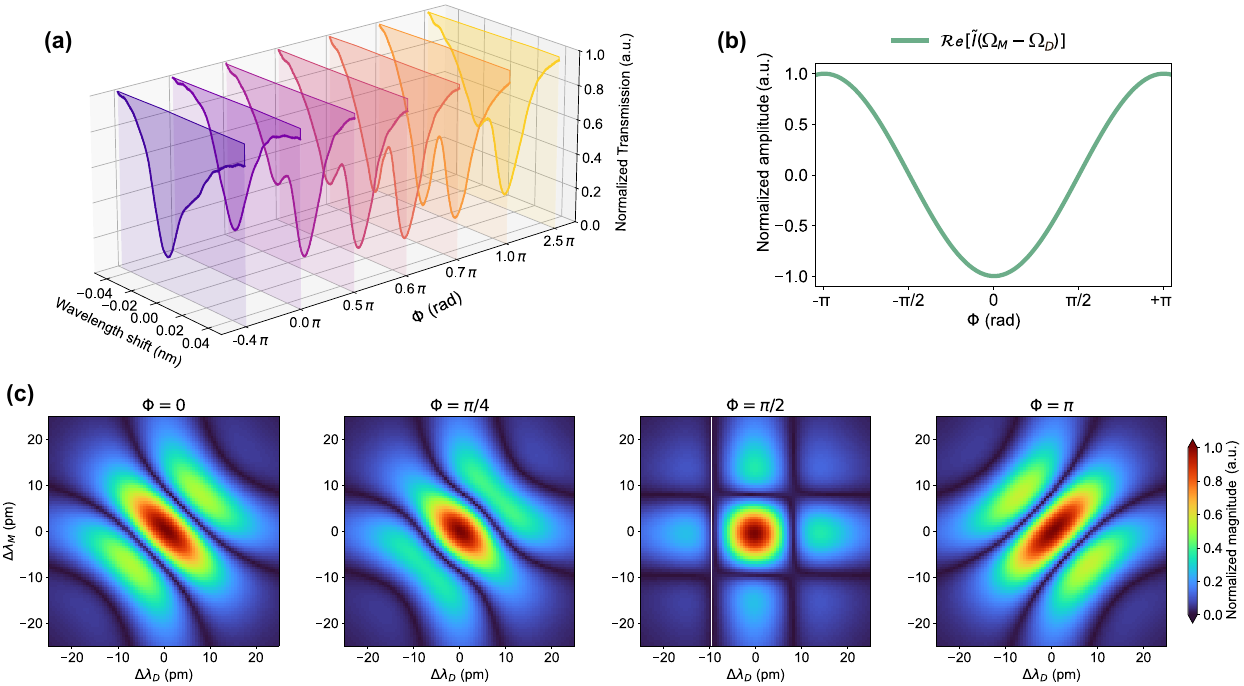}  
\captionof{figure}{\textbf{Additional WS characterizations and simulations.} (a) Experimental spectra of a single WS unit varying the applied phase shift $\Phi$. (b) Simulated real part of the Fourier components at $\Omega_M - \Omega_D$ as a function of the phase-shift $\Phi$ applied by the WS-unit. The shown amplitude is normalized to the maximum of the curve. (c) Simulated magnitude of the Fourier component at $2(\Omega_D + \Omega_M)$ for for different phases. For (b) and (c), the corresponding experimental results are reported in the main text.}
\label{fig:WS_cal}
\end{center}
\end{widetext}

\subsection*{Quantum walk simulation}
The quantum walks shown in Fig.\ref{fig:Results_Q}(b) of the main text are also reported in Fig.~\ref{fig:Results_QW} to enable direct comparison with the simulations. The state at the input of OUT PM is written as $\ket{\Psi}=\frac{1}{\mathcal{N}}\sum_{l=0}^{5}\beta_l\ket{ll}_{si}$, where $\mathcal{N}$ is a normalization constant. The four frequency bins labeled from 1 to 4 span the $4\times4$ Hilbert space of interest, within which frequency-resolved coincidence measurements are performed. The outer bins, labeled 0 and 5, also participate in the quantum walk but cannot be measured due to the limited bandwidth ($\sim80\,\mathrm{GHz}$) of the off-chip filters used for pump suppression. Frequency bins with indices $l<0$ or $l>5$ are not included in the simulations owing to the relatively small modulation depth ($\delta\sim0.8\,$rad) applied to the OUT PM, which confines frequency-bin scrambling primarily to nearest neighbors.
The coefficients $|\beta_l|^2$ for $l=1$–$4$ are extracted directly from the measured JSI of the source. The dominant source of inhomogeneity among the coefficients—visible as a decreasing intensity along the main diagonal in the initial JSI panel—is the pump-rejection filter, which imposes a sinusoidal envelope with a free spectral range of $500\,\mathrm{GHz}$ on the broadband frequency comb. This envelope is used to extrapolate the values of $|\beta_0|$ and $|\beta_5|$.\\
To extract the relative phases between the input frequency bins, we sent the state $\ket{\Psi}$ through the OUT PM with the modulation depth set to $\delta=0.8\,$rad, and measured the resulting JSI, shown in Fig.~\ref{fig:Results_QW}(b). We then employed a numerical optimization procedure (\texttt{scipy.minimize}) to maximize the overlap between the experimentally measured JSI and a simulated distribution. The quantum-walk dynamics were simulated following the method described in the Supplemental document of Ref.~\cite{2020_Imany_QWalk}, which we do not reproduce here for brevity. The free parameters in the optimization are the relative phases $\mathrm{Arg}(\beta_l)$ for $l=1-5$, with $\mathrm{Arg}(\beta_0)=0$ used as a reference. All other parameters were fixed based on independent experimental calibration, including the modulation depth $\delta$, the losses of each WS channel, and the applied spectral phases $\{\Phi_l\}_{l=1}^{4}$, which were all set to zero. We find that the phases $\mathrm{Arg}(\beta_l)$ maximizing the overlap between the simulated ($\textrm{JSI}_{\textup{t}}$) and the experimental ($\textrm{JSI}_{\textup{e}}$) JSIs are all distributed within $\pm 0.1$ rad. The simulation, shown in Fig.~\ref{fig:Results_QW}(b), has a fidelity $\mathcal{F}=\textrm{Tr}([\textrm{JSI}_{\textup{t}}]^{\dagger}\textrm{JSI}_{\textup{e}})$ of $\mathcal{F}=99.38(5)\%$.  This indicates that the input frequency bins are mutually phase aligned within experimental and numerical uncertainty. Consequently, the generated photon pairs undergo a correlated quantum walk, characterized by frequency correlations spreading away from the main diagonal of the JSI \cite{2020_Imany_QWalk}.
To convert this behavior into an  anti-correlated quantum walk, in which frequency correlations remain confined along the main diagonal, the input state must be transformed into $\ket{\Psi}_{\textup{ac}}=\sum_{l=0}^{5}|\beta_l|e^{il\pi}$, thus a relative $\pi$ phase shift must be introduced between consecutive frequency bins. We achieve this transformation using the four-channel WS. Specifically, we apply the phase pattern  $\Phi_{1-4}=[-0.5\pi,0.5\pi,-0.5\pi,0.5\pi]$. The experimentally measured JSI corresponding to this configuration is shown in Fig.~\ref{fig:Results_QW}(c). After the OUT PM, the frequency correlations remain confined along the main diagonal, clearly indicating the transition to an anti-correlated quantum walk. The fidelity between the experimental JSI and the corresponding numerical simulation reported in Fig.~\ref{fig:Results_QW}(c) is $\mathcal{F}=98.7(5) \%$.
\begin{widetext}
\begin{center}
\includegraphics[width = \textwidth]{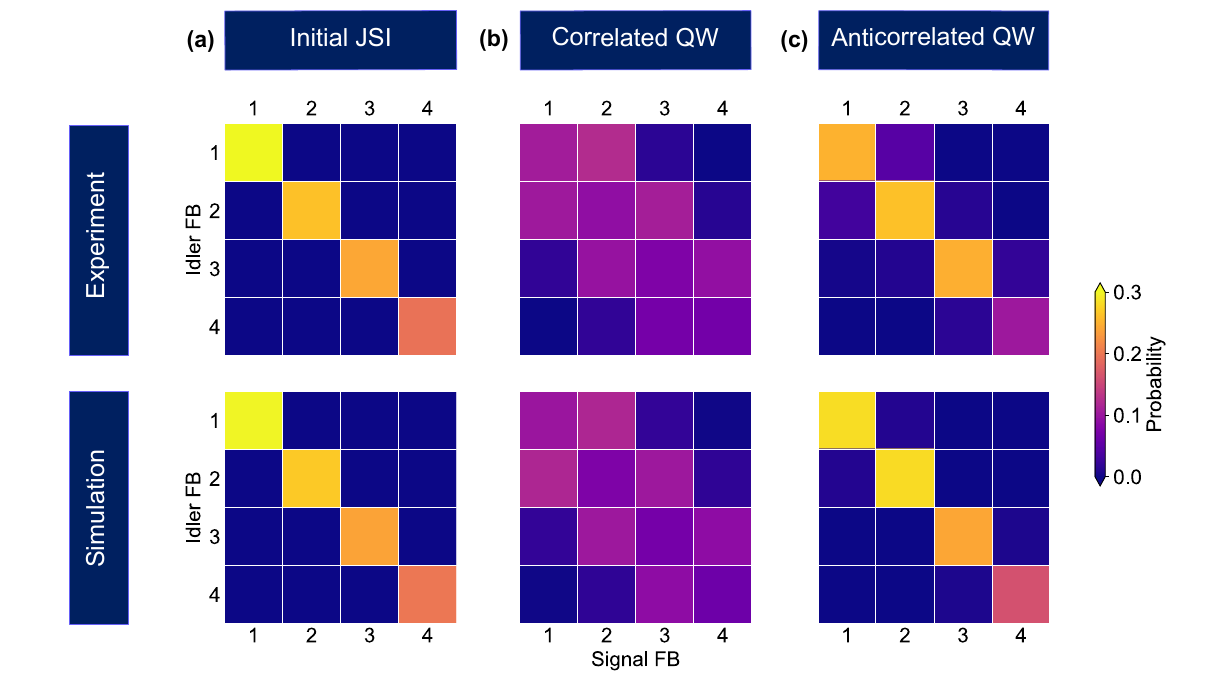}
\captionof{figure}{\textbf{Experimental and simulated quantum walks.} (a) Initial JSI, coincidence pattern obtained directly after the source, keeping the PMs off and all the WS-units in the PASS configuration. It corresponds to the state $\ket{\Psi}$ described in the text.
(b) Measured JSI after the OUT PM, showing a correlated quantum walk.
(c) JSI measured after the OUT PM when the input state is $\ket{\Psi}_{\textup{ac}}$ (see main text for details), resulting in a anti-correlated quantum walk with low energy leakage outside the main diagonal.
Top panels show the JSIs measured experimentally via frequency-resolved coincidence measurements. Bottom panels show the corresponding JSIs simulated using the model described in the Supplemental Document of Ref.~\cite{2020_Imany_QWalk}. Both experimental and simulated matrix entries have been normalized to the integral over the full $4\times 4$ space.}
\label{fig:Results_QW}
\end{center}
\end{widetext}


\begin{thebibliography}{67}%
\makeatletter
\providecommand \@ifxundefined [1]{%
 \@ifx{#1\undefined}
}%
\providecommand \@ifnum [1]{%
 \ifnum #1\expandafter \@firstoftwo
 \else \expandafter \@secondoftwo
 \fi
}%
\providecommand \@ifx [1]{%
 \ifx #1\expandafter \@firstoftwo
 \else \expandafter \@secondoftwo
 \fi
}%
\providecommand \natexlab [1]{#1}%
\providecommand \enquote  [1]{``#1''}%
\providecommand \bibnamefont  [1]{#1}%
\providecommand \bibfnamefont [1]{#1}%
\providecommand \citenamefont [1]{#1}%
\providecommand \href@noop [0]{\@secondoftwo}%
\providecommand \href [0]{\begingroup \@sanitize@url \@href}%
\providecommand \@href[1]{\@@startlink{#1}\@@href}%
\providecommand \@@href[1]{\endgroup#1\@@endlink}%
\providecommand \@sanitize@url [0]{\catcode `\\12\catcode `\$12\catcode
  `\&12\catcode `\#12\catcode `\^12\catcode `\_12\catcode `\%12\relax}%
\providecommand \@@startlink[1]{}%
\providecommand \@@endlink[0]{}%
\providecommand \url  [0]{\begingroup\@sanitize@url \@url }%
\providecommand \@url [1]{\endgroup\@href {#1}{\urlprefix }}%
\providecommand \urlprefix  [0]{URL }%
\providecommand \Eprint [0]{\href }%
\providecommand \doibase [0]{https://doi.org/}%
\providecommand \selectlanguage [0]{\@gobble}%
\providecommand \bibinfo  [0]{\@secondoftwo}%
\providecommand \bibfield  [0]{\@secondoftwo}%
\providecommand \translation [1]{[#1]}%
\providecommand \BibitemOpen [0]{}%
\providecommand \bibitemStop [0]{}%
\providecommand \bibitemNoStop [0]{.\EOS\space}%
\providecommand \EOS [0]{\spacefactor3000\relax}%
\providecommand \BibitemShut  [1]{\csname bibitem#1\endcsname}%
\let\auto@bib@innerbib\@empty
\bibitem [{\citenamefont {Shi}\ \emph {et~al.}(2020)\citenamefont {Shi},
  \citenamefont {Tian},\ and\ \citenamefont {Gervais}}]{shi2020scaling}%
  \BibitemOpen
  \bibfield  {author} {\bibinfo {author} {\bibfnamefont {W.}~\bibnamefont
  {Shi}}, \bibinfo {author} {\bibfnamefont {Y.}~\bibnamefont {Tian}},\ and\
  \bibinfo {author} {\bibfnamefont {A.}~\bibnamefont {Gervais}},\ }\bibfield
  {title} {\bibinfo {title} {Scaling capacity of fiber-optic transmission
  systems via silicon photonics},\ }\href@noop {} {\bibfield  {journal}
  {\bibinfo  {journal} {Nanophotonics}\ }\textbf {\bibinfo {volume} {9}},\
  \bibinfo {pages} {4629} (\bibinfo {year} {2020})}\BibitemShut {NoStop}%
\bibitem [{\citenamefont {Xie}\ \emph {et~al.}(2022)\citenamefont {Xie},
  \citenamefont {Hong}, \citenamefont {Yan}, \citenamefont {Zhang},
  \citenamefont {Zhang}, \citenamefont {Zhuang},\ and\ \citenamefont
  {Dai}}]{xie2022low}%
  \BibitemOpen
  \bibfield  {author} {\bibinfo {author} {\bibfnamefont {Y.}~\bibnamefont
  {Xie}}, \bibinfo {author} {\bibfnamefont {S.}~\bibnamefont {Hong}}, \bibinfo
  {author} {\bibfnamefont {H.}~\bibnamefont {Yan}}, \bibinfo {author}
  {\bibfnamefont {C.}~\bibnamefont {Zhang}}, \bibinfo {author} {\bibfnamefont
  {L.}~\bibnamefont {Zhang}}, \bibinfo {author} {\bibfnamefont
  {L.}~\bibnamefont {Zhuang}},\ and\ \bibinfo {author} {\bibfnamefont
  {D.}~\bibnamefont {Dai}},\ }\bibfield  {title} {\bibinfo {title} {Low-loss
  chip-scale programmable silicon photonic processor},\ }\href@noop {}
  {\bibfield  {journal} {\bibinfo  {journal} {Opto-Electronic Advances}\
  }\textbf {\bibinfo {volume} {6}},\ \bibinfo {pages} {220030} (\bibinfo {year}
  {2022})}\BibitemShut {NoStop}%
\bibitem [{\citenamefont {Han}\ \emph {et~al.}(2025{\natexlab{a}})\citenamefont
  {Han}, \citenamefont {Yang}, \citenamefont {Qin}, \citenamefont {Zhou},
  \citenamefont {Zheng}, \citenamefont {Zhang}, \citenamefont {Wang},
  \citenamefont {Sun}, \citenamefont {Lu}, \citenamefont {Wang} \emph
  {et~al.}}]{han2025exploring}%
  \BibitemOpen
  \bibfield  {author} {\bibinfo {author} {\bibfnamefont {C.}~\bibnamefont
  {Han}}, \bibinfo {author} {\bibfnamefont {Q.}~\bibnamefont {Yang}}, \bibinfo
  {author} {\bibfnamefont {J.}~\bibnamefont {Qin}}, \bibinfo {author}
  {\bibfnamefont {Y.}~\bibnamefont {Zhou}}, \bibinfo {author} {\bibfnamefont
  {Z.}~\bibnamefont {Zheng}}, \bibinfo {author} {\bibfnamefont
  {Y.}~\bibnamefont {Zhang}}, \bibinfo {author} {\bibfnamefont
  {H.}~\bibnamefont {Wang}}, \bibinfo {author} {\bibfnamefont {Y.}~\bibnamefont
  {Sun}}, \bibinfo {author} {\bibfnamefont {J.}~\bibnamefont {Lu}}, \bibinfo
  {author} {\bibfnamefont {Y.}~\bibnamefont {Wang}}, \emph {et~al.},\
  }\bibfield  {title} {\bibinfo {title} {Exploring 400 gbps/$\lambda$ and
  beyond with ai-accelerated silicon photonic slow-light technology},\
  }\href@noop {} {\bibfield  {journal} {\bibinfo  {journal} {Nature
  Communications}\ }\textbf {\bibinfo {volume} {16}},\ \bibinfo {pages} {6547}
  (\bibinfo {year} {2025}{\natexlab{a}})}\BibitemShut {NoStop}%
\bibitem [{\citenamefont {Deng}\ \emph {et~al.}(2025)\citenamefont {Deng},
  \citenamefont {Zhang}, \citenamefont {Soltanian}, \citenamefont {Chen},
  \citenamefont {Pang}, \citenamefont {Vaissiere}, \citenamefont {Neel},
  \citenamefont {Ramirez}, \citenamefont {Decobert}, \citenamefont {Singh}
  \emph {et~al.}}]{deng2025single}%
  \BibitemOpen
  \bibfield  {author} {\bibinfo {author} {\bibfnamefont {H.}~\bibnamefont
  {Deng}}, \bibinfo {author} {\bibfnamefont {J.}~\bibnamefont {Zhang}},
  \bibinfo {author} {\bibfnamefont {E.}~\bibnamefont {Soltanian}}, \bibinfo
  {author} {\bibfnamefont {X.}~\bibnamefont {Chen}}, \bibinfo {author}
  {\bibfnamefont {C.}~\bibnamefont {Pang}}, \bibinfo {author} {\bibfnamefont
  {N.}~\bibnamefont {Vaissiere}}, \bibinfo {author} {\bibfnamefont
  {D.}~\bibnamefont {Neel}}, \bibinfo {author} {\bibfnamefont {J.}~\bibnamefont
  {Ramirez}}, \bibinfo {author} {\bibfnamefont {J.}~\bibnamefont {Decobert}},
  \bibinfo {author} {\bibfnamefont {N.}~\bibnamefont {Singh}}, \emph {et~al.},\
  }\bibfield  {title} {\bibinfo {title} {Single-chip silicon photonic engine
  for analog optical and microwave signals processing},\ }\href@noop {}
  {\bibfield  {journal} {\bibinfo  {journal} {Nature Communications}\ }\textbf
  {\bibinfo {volume} {16}},\ \bibinfo {pages} {5087} (\bibinfo {year}
  {2025})}\BibitemShut {NoStop}%
\bibitem [{\citenamefont {Liu}\ \emph {et~al.}(2022)\citenamefont {Liu},
  \citenamefont {Wang}, \citenamefont {Wang}, \citenamefont {Li}, \citenamefont
  {Yu},\ and\ \citenamefont {Cai}}]{liu2022silicon}%
  \BibitemOpen
  \bibfield  {author} {\bibinfo {author} {\bibfnamefont {Y.}~\bibnamefont
  {Liu}}, \bibinfo {author} {\bibfnamefont {S.}~\bibnamefont {Wang}}, \bibinfo
  {author} {\bibfnamefont {J.}~\bibnamefont {Wang}}, \bibinfo {author}
  {\bibfnamefont {X.}~\bibnamefont {Li}}, \bibinfo {author} {\bibfnamefont
  {M.}~\bibnamefont {Yu}},\ and\ \bibinfo {author} {\bibfnamefont
  {Y.}~\bibnamefont {Cai}},\ }\bibfield  {title} {\bibinfo {title} {Silicon
  photonic transceivers in the field of optical communication},\ }\href@noop {}
  {\bibfield  {journal} {\bibinfo  {journal} {Nano communication networks}\
  }\textbf {\bibinfo {volume} {31}},\ \bibinfo {pages} {100379} (\bibinfo
  {year} {2022})}\BibitemShut {NoStop}%
\bibitem [{\citenamefont {Wang}\ \emph {et~al.}(2015)\citenamefont {Wang},
  \citenamefont {Shen}, \citenamefont {Fan}, \citenamefont {Wu}, \citenamefont
  {Niu}, \citenamefont {Varghese}, \citenamefont {Xuan}, \citenamefont
  {Leaird}, \citenamefont {Wang}, \citenamefont {Gan} \emph
  {et~al.}}]{wang2015reconfigurable}%
  \BibitemOpen
  \bibfield  {author} {\bibinfo {author} {\bibfnamefont {J.}~\bibnamefont
  {Wang}}, \bibinfo {author} {\bibfnamefont {H.}~\bibnamefont {Shen}}, \bibinfo
  {author} {\bibfnamefont {L.}~\bibnamefont {Fan}}, \bibinfo {author}
  {\bibfnamefont {R.}~\bibnamefont {Wu}}, \bibinfo {author} {\bibfnamefont
  {B.}~\bibnamefont {Niu}}, \bibinfo {author} {\bibfnamefont {L.~T.}\
  \bibnamefont {Varghese}}, \bibinfo {author} {\bibfnamefont {Y.}~\bibnamefont
  {Xuan}}, \bibinfo {author} {\bibfnamefont {D.~E.}\ \bibnamefont {Leaird}},
  \bibinfo {author} {\bibfnamefont {X.}~\bibnamefont {Wang}}, \bibinfo {author}
  {\bibfnamefont {F.}~\bibnamefont {Gan}}, \emph {et~al.},\ }\bibfield  {title}
  {\bibinfo {title} {Reconfigurable radio-frequency arbitrary waveforms
  synthesized in a silicon photonic chip},\ }\href@noop {} {\bibfield
  {journal} {\bibinfo  {journal} {Nature communications}\ }\textbf {\bibinfo
  {volume} {6}},\ \bibinfo {pages} {5957} (\bibinfo {year} {2015})}\BibitemShut
  {NoStop}%
\bibitem [{\citenamefont {Tao}\ \emph {et~al.}(2022)\citenamefont {Tao},
  \citenamefont {Yang}, \citenamefont {Tao}, \citenamefont {Chang},
  \citenamefont {Shu}, \citenamefont {Jin}, \citenamefont {Zhou}, \citenamefont
  {Ge},\ and\ \citenamefont {Wang}}]{tao2022fully}%
  \BibitemOpen
  \bibfield  {author} {\bibinfo {author} {\bibfnamefont {Y.}~\bibnamefont
  {Tao}}, \bibinfo {author} {\bibfnamefont {F.}~\bibnamefont {Yang}}, \bibinfo
  {author} {\bibfnamefont {Z.}~\bibnamefont {Tao}}, \bibinfo {author}
  {\bibfnamefont {L.}~\bibnamefont {Chang}}, \bibinfo {author} {\bibfnamefont
  {H.}~\bibnamefont {Shu}}, \bibinfo {author} {\bibfnamefont {M.}~\bibnamefont
  {Jin}}, \bibinfo {author} {\bibfnamefont {Y.}~\bibnamefont {Zhou}}, \bibinfo
  {author} {\bibfnamefont {Z.}~\bibnamefont {Ge}},\ and\ \bibinfo {author}
  {\bibfnamefont {X.}~\bibnamefont {Wang}},\ }\bibfield  {title} {\bibinfo
  {title} {Fully on-chip microwave photonic instantaneous frequency measurement
  system},\ }\href@noop {} {\bibfield  {journal} {\bibinfo  {journal} {Laser \&
  Photonics Reviews}\ }\textbf {\bibinfo {volume} {16}},\ \bibinfo {pages}
  {2200158} (\bibinfo {year} {2022})}\BibitemShut {NoStop}%
\bibitem [{\citenamefont {Bao}\ \emph {et~al.}(2023)\citenamefont {Bao},
  \citenamefont {Fu}, \citenamefont {Pramanik}, \citenamefont {Mao},
  \citenamefont {Chi}, \citenamefont {Cao}, \citenamefont {Zhai}, \citenamefont
  {Mao}, \citenamefont {Dai}, \citenamefont {Chen} \emph
  {et~al.}}]{bao2023very}%
  \BibitemOpen
  \bibfield  {author} {\bibinfo {author} {\bibfnamefont {J.}~\bibnamefont
  {Bao}}, \bibinfo {author} {\bibfnamefont {Z.}~\bibnamefont {Fu}}, \bibinfo
  {author} {\bibfnamefont {T.}~\bibnamefont {Pramanik}}, \bibinfo {author}
  {\bibfnamefont {J.}~\bibnamefont {Mao}}, \bibinfo {author} {\bibfnamefont
  {Y.}~\bibnamefont {Chi}}, \bibinfo {author} {\bibfnamefont {Y.}~\bibnamefont
  {Cao}}, \bibinfo {author} {\bibfnamefont {C.}~\bibnamefont {Zhai}}, \bibinfo
  {author} {\bibfnamefont {Y.}~\bibnamefont {Mao}}, \bibinfo {author}
  {\bibfnamefont {T.}~\bibnamefont {Dai}}, \bibinfo {author} {\bibfnamefont
  {X.}~\bibnamefont {Chen}}, \emph {et~al.},\ }\bibfield  {title} {\bibinfo
  {title} {Very-large-scale integrated quantum graph photonics},\ }\href@noop
  {} {\bibfield  {journal} {\bibinfo  {journal} {Nature Photonics}\ }\textbf
  {\bibinfo {volume} {17}},\ \bibinfo {pages} {573} (\bibinfo {year}
  {2023})}\BibitemShut {NoStop}%
\bibitem [{\citenamefont {Qiang}\ \emph {et~al.}(2018)\citenamefont {Qiang},
  \citenamefont {Zhou}, \citenamefont {Wang}, \citenamefont {Wilkes},
  \citenamefont {Loke}, \citenamefont {O’Gara}, \citenamefont {Kling},
  \citenamefont {Marshall}, \citenamefont {Santagati}, \citenamefont {Ralph}
  \emph {et~al.}}]{qiang2018large}%
  \BibitemOpen
  \bibfield  {author} {\bibinfo {author} {\bibfnamefont {X.}~\bibnamefont
  {Qiang}}, \bibinfo {author} {\bibfnamefont {X.}~\bibnamefont {Zhou}},
  \bibinfo {author} {\bibfnamefont {J.}~\bibnamefont {Wang}}, \bibinfo {author}
  {\bibfnamefont {C.~M.}\ \bibnamefont {Wilkes}}, \bibinfo {author}
  {\bibfnamefont {T.}~\bibnamefont {Loke}}, \bibinfo {author} {\bibfnamefont
  {S.}~\bibnamefont {O’Gara}}, \bibinfo {author} {\bibfnamefont
  {L.}~\bibnamefont {Kling}}, \bibinfo {author} {\bibfnamefont {G.~D.}\
  \bibnamefont {Marshall}}, \bibinfo {author} {\bibfnamefont {R.}~\bibnamefont
  {Santagati}}, \bibinfo {author} {\bibfnamefont {T.~C.}\ \bibnamefont
  {Ralph}}, \emph {et~al.},\ }\bibfield  {title} {\bibinfo {title} {Large-scale
  silicon quantum photonics implementing arbitrary two-qubit processing},\
  }\href@noop {} {\bibfield  {journal} {\bibinfo  {journal} {Nature photonics}\
  }\textbf {\bibinfo {volume} {12}},\ \bibinfo {pages} {534} (\bibinfo {year}
  {2018})}\BibitemShut {NoStop}%
\bibitem [{\citenamefont {Maring}\ \emph {et~al.}(2024)\citenamefont {Maring},
  \citenamefont {Fyrillas}, \citenamefont {Pont}, \citenamefont {Ivanov},
  \citenamefont {Stepanov}, \citenamefont {Margaria}, \citenamefont {Hease},
  \citenamefont {Pishchagin}, \citenamefont {Lema{\^\i}tre}, \citenamefont
  {Sagnes} \emph {et~al.}}]{maring2024versatile}%
  \BibitemOpen
  \bibfield  {author} {\bibinfo {author} {\bibfnamefont {N.}~\bibnamefont
  {Maring}}, \bibinfo {author} {\bibfnamefont {A.}~\bibnamefont {Fyrillas}},
  \bibinfo {author} {\bibfnamefont {M.}~\bibnamefont {Pont}}, \bibinfo {author}
  {\bibfnamefont {E.}~\bibnamefont {Ivanov}}, \bibinfo {author} {\bibfnamefont
  {P.}~\bibnamefont {Stepanov}}, \bibinfo {author} {\bibfnamefont
  {N.}~\bibnamefont {Margaria}}, \bibinfo {author} {\bibfnamefont
  {W.}~\bibnamefont {Hease}}, \bibinfo {author} {\bibfnamefont
  {A.}~\bibnamefont {Pishchagin}}, \bibinfo {author} {\bibfnamefont
  {A.}~\bibnamefont {Lema{\^\i}tre}}, \bibinfo {author} {\bibfnamefont
  {I.}~\bibnamefont {Sagnes}}, \emph {et~al.},\ }\bibfield  {title} {\bibinfo
  {title} {A versatile single-photon-based quantum computing platform},\
  }\href@noop {} {\bibfield  {journal} {\bibinfo  {journal} {Nature Photonics}\
  }\textbf {\bibinfo {volume} {18}},\ \bibinfo {pages} {603} (\bibinfo {year}
  {2024})}\BibitemShut {NoStop}%
\bibitem [{\citenamefont {Vigliar}\ \emph {et~al.}(2021)\citenamefont
  {Vigliar}, \citenamefont {Paesani}, \citenamefont {Ding}, \citenamefont
  {Adcock}, \citenamefont {Wang}, \citenamefont {Morley-Short}, \citenamefont
  {Bacco}, \citenamefont {Oxenl{\o}we}, \citenamefont {Thompson}, \citenamefont
  {Rarity} \emph {et~al.}}]{vigliar2021error}%
  \BibitemOpen
  \bibfield  {author} {\bibinfo {author} {\bibfnamefont {C.}~\bibnamefont
  {Vigliar}}, \bibinfo {author} {\bibfnamefont {S.}~\bibnamefont {Paesani}},
  \bibinfo {author} {\bibfnamefont {Y.}~\bibnamefont {Ding}}, \bibinfo {author}
  {\bibfnamefont {J.~C.}\ \bibnamefont {Adcock}}, \bibinfo {author}
  {\bibfnamefont {J.}~\bibnamefont {Wang}}, \bibinfo {author} {\bibfnamefont
  {S.}~\bibnamefont {Morley-Short}}, \bibinfo {author} {\bibfnamefont
  {D.}~\bibnamefont {Bacco}}, \bibinfo {author} {\bibfnamefont {L.~K.}\
  \bibnamefont {Oxenl{\o}we}}, \bibinfo {author} {\bibfnamefont {M.~G.}\
  \bibnamefont {Thompson}}, \bibinfo {author} {\bibfnamefont {J.~G.}\
  \bibnamefont {Rarity}}, \emph {et~al.},\ }\bibfield  {title} {\bibinfo
  {title} {Error-protected qubits in a silicon photonic chip},\ }\href@noop {}
  {\bibfield  {journal} {\bibinfo  {journal} {Nature Physics}\ }\textbf
  {\bibinfo {volume} {17}},\ \bibinfo {pages} {1137} (\bibinfo {year}
  {2021})}\BibitemShut {NoStop}%
\bibitem [{\citenamefont {Finco}\ \emph {et~al.}(2024)\citenamefont {Finco},
  \citenamefont {Miserocchi}, \citenamefont {Maeder}, \citenamefont {Kellner},
  \citenamefont {Sabatti}, \citenamefont {Chapman},\ and\ \citenamefont
  {Grange}}]{finco2024time}%
  \BibitemOpen
  \bibfield  {author} {\bibinfo {author} {\bibfnamefont {G.}~\bibnamefont
  {Finco}}, \bibinfo {author} {\bibfnamefont {F.}~\bibnamefont {Miserocchi}},
  \bibinfo {author} {\bibfnamefont {A.}~\bibnamefont {Maeder}}, \bibinfo
  {author} {\bibfnamefont {J.}~\bibnamefont {Kellner}}, \bibinfo {author}
  {\bibfnamefont {A.}~\bibnamefont {Sabatti}}, \bibinfo {author} {\bibfnamefont
  {R.~J.}\ \bibnamefont {Chapman}},\ and\ \bibinfo {author} {\bibfnamefont
  {R.}~\bibnamefont {Grange}},\ }\bibfield  {title} {\bibinfo {title} {Time-bin
  entangled bell state generation and tomography on thin-film lithium
  niobate},\ }\href@noop {} {\bibfield  {journal} {\bibinfo  {journal} {npj
  Quantum Information}\ }\textbf {\bibinfo {volume} {10}},\ \bibinfo {pages}
  {135} (\bibinfo {year} {2024})}\BibitemShut {NoStop}%
\bibitem [{\citenamefont {Yu}\ \emph {et~al.}(2025)\citenamefont {Yu},
  \citenamefont {Sciara}, \citenamefont {Chemnitz}, \citenamefont {Montaut},
  \citenamefont {Crockett}, \citenamefont {Fischer}, \citenamefont {Helsten},
  \citenamefont {Wetzel}, \citenamefont {Goebel}, \citenamefont {Kr{\"a}mer}
  \emph {et~al.}}]{yu2025quantum}%
  \BibitemOpen
  \bibfield  {author} {\bibinfo {author} {\bibfnamefont {H.}~\bibnamefont
  {Yu}}, \bibinfo {author} {\bibfnamefont {S.}~\bibnamefont {Sciara}}, \bibinfo
  {author} {\bibfnamefont {M.}~\bibnamefont {Chemnitz}}, \bibinfo {author}
  {\bibfnamefont {N.}~\bibnamefont {Montaut}}, \bibinfo {author} {\bibfnamefont
  {B.}~\bibnamefont {Crockett}}, \bibinfo {author} {\bibfnamefont
  {B.}~\bibnamefont {Fischer}}, \bibinfo {author} {\bibfnamefont
  {R.}~\bibnamefont {Helsten}}, \bibinfo {author} {\bibfnamefont
  {B.}~\bibnamefont {Wetzel}}, \bibinfo {author} {\bibfnamefont {T.~A.}\
  \bibnamefont {Goebel}}, \bibinfo {author} {\bibfnamefont {R.~G.}\
  \bibnamefont {Kr{\"a}mer}}, \emph {et~al.},\ }\bibfield  {title} {\bibinfo
  {title} {Quantum key distribution implemented with d-level time-bin entangled
  photons},\ }\href@noop {} {\bibfield  {journal} {\bibinfo  {journal} {Nature
  Communications}\ }\textbf {\bibinfo {volume} {16}},\ \bibinfo {pages} {171}
  (\bibinfo {year} {2025})}\BibitemShut {NoStop}%
\bibitem [{\citenamefont {Feng}\ \emph {et~al.}(2022)\citenamefont {Feng},
  \citenamefont {Zhang}, \citenamefont {Xiong}, \citenamefont {Liu},
  \citenamefont {Cheng}, \citenamefont {Jing}, \citenamefont {Qi},
  \citenamefont {Chen}, \citenamefont {He}, \citenamefont {Guo} \emph
  {et~al.}}]{feng2022transverse}%
  \BibitemOpen
  \bibfield  {author} {\bibinfo {author} {\bibfnamefont {L.-T.}\ \bibnamefont
  {Feng}}, \bibinfo {author} {\bibfnamefont {M.}~\bibnamefont {Zhang}},
  \bibinfo {author} {\bibfnamefont {X.}~\bibnamefont {Xiong}}, \bibinfo
  {author} {\bibfnamefont {D.}~\bibnamefont {Liu}}, \bibinfo {author}
  {\bibfnamefont {Y.-J.}\ \bibnamefont {Cheng}}, \bibinfo {author}
  {\bibfnamefont {F.-M.}\ \bibnamefont {Jing}}, \bibinfo {author}
  {\bibfnamefont {X.-Z.}\ \bibnamefont {Qi}}, \bibinfo {author} {\bibfnamefont
  {Y.}~\bibnamefont {Chen}}, \bibinfo {author} {\bibfnamefont {D.-Y.}\
  \bibnamefont {He}}, \bibinfo {author} {\bibfnamefont {G.-P.}\ \bibnamefont
  {Guo}}, \emph {et~al.},\ }\bibfield  {title} {\bibinfo {title} {Transverse
  mode-encoded quantum gate on a silicon photonic chip},\ }\href@noop {}
  {\bibfield  {journal} {\bibinfo  {journal} {Physical Review Letters}\
  }\textbf {\bibinfo {volume} {128}},\ \bibinfo {pages} {060501} (\bibinfo
  {year} {2022})}\BibitemShut {NoStop}%
\bibitem [{\citenamefont {Forbes}\ \emph {et~al.}(2025)\citenamefont {Forbes},
  \citenamefont {Yard}, \citenamefont {Bielak}, \citenamefont {Thomas},
  \citenamefont {Jones}, \citenamefont {Paesani}, \citenamefont {Borghi},\ and\
  \citenamefont {Laing}}]{forbes2025hybrid}%
  \BibitemOpen
  \bibfield  {author} {\bibinfo {author} {\bibfnamefont {I.}~\bibnamefont
  {Forbes}}, \bibinfo {author} {\bibfnamefont {P.}~\bibnamefont {Yard}},
  \bibinfo {author} {\bibfnamefont {M.}~\bibnamefont {Bielak}}, \bibinfo
  {author} {\bibfnamefont {M.~A.}\ \bibnamefont {Thomas}}, \bibinfo {author}
  {\bibfnamefont {M.~S.}\ \bibnamefont {Jones}}, \bibinfo {author}
  {\bibfnamefont {S.}~\bibnamefont {Paesani}}, \bibinfo {author} {\bibfnamefont
  {M.}~\bibnamefont {Borghi}},\ and\ \bibinfo {author} {\bibfnamefont
  {A.}~\bibnamefont {Laing}},\ }\bibfield  {title} {\bibinfo {title} {Hybrid
  path-transverse electric mode qudit encoding on an integrated photonic
  chip},\ }\href@noop {} {\bibfield  {journal} {\bibinfo  {journal} {arXiv
  preprint arXiv:2510.15774}\ } (\bibinfo {year} {2025})}\BibitemShut {NoStop}%
\bibitem [{\citenamefont {Zhang}\ \emph {et~al.}(2024)\citenamefont {Zhang},
  \citenamefont {Wu},\ and\ \citenamefont {Poon}}]{zhang2024polarization}%
  \BibitemOpen
  \bibfield  {author} {\bibinfo {author} {\bibfnamefont {Q.}~\bibnamefont
  {Zhang}}, \bibinfo {author} {\bibfnamefont {K.}~\bibnamefont {Wu}},\ and\
  \bibinfo {author} {\bibfnamefont {A.~W.}\ \bibnamefont {Poon}},\ }\bibfield
  {title} {\bibinfo {title} {Polarization entanglement generation in silicon
  nitride waveguide-coupled dual microring resonators},\ }\href@noop {}
  {\bibfield  {journal} {\bibinfo  {journal} {Optics Express}\ }\textbf
  {\bibinfo {volume} {32}},\ \bibinfo {pages} {22804} (\bibinfo {year}
  {2024})}\BibitemShut {NoStop}%
\bibitem [{\citenamefont {Miloshevsky}\ \emph {et~al.}(2024)\citenamefont
  {Miloshevsky}, \citenamefont {Cohen}, \citenamefont {Myilswamy},
  \citenamefont {Alshowkan}, \citenamefont {Fatema}, \citenamefont {Lu},
  \citenamefont {Weiner},\ and\ \citenamefont {Lukens}}]{miloshevsky2024cmos}%
  \BibitemOpen
  \bibfield  {author} {\bibinfo {author} {\bibfnamefont {A.}~\bibnamefont
  {Miloshevsky}}, \bibinfo {author} {\bibfnamefont {L.~M.}\ \bibnamefont
  {Cohen}}, \bibinfo {author} {\bibfnamefont {K.~V.}\ \bibnamefont
  {Myilswamy}}, \bibinfo {author} {\bibfnamefont {M.}~\bibnamefont
  {Alshowkan}}, \bibinfo {author} {\bibfnamefont {S.}~\bibnamefont {Fatema}},
  \bibinfo {author} {\bibfnamefont {H.-H.}\ \bibnamefont {Lu}}, \bibinfo
  {author} {\bibfnamefont {A.~M.}\ \bibnamefont {Weiner}},\ and\ \bibinfo
  {author} {\bibfnamefont {J.~M.}\ \bibnamefont {Lukens}},\ }\bibfield  {title}
  {\bibinfo {title} {Cmos photonic integrated source of broadband
  polarization-entangled photons},\ }\href@noop {} {\bibfield  {journal}
  {\bibinfo  {journal} {Optica Quantum}\ }\textbf {\bibinfo {volume} {2}},\
  \bibinfo {pages} {254} (\bibinfo {year} {2024})}\BibitemShut {NoStop}%
\bibitem [{\citenamefont {Lukens}\ and\ \citenamefont
  {Lougovski}(2016)}]{2016_Lukens_QFP_theory}%
  \BibitemOpen
  \bibfield  {author} {\bibinfo {author} {\bibfnamefont {J.~M.}\ \bibnamefont
  {Lukens}}\ and\ \bibinfo {author} {\bibfnamefont {P.}~\bibnamefont
  {Lougovski}},\ }\bibfield  {title} {\bibinfo {title} {Frequency-encoded
  photonic qubits for scalable quantum information processing},\ }\href@noop {}
  {\bibfield  {journal} {\bibinfo  {journal} {Optica}\ }\textbf {\bibinfo
  {volume} {4}},\ \bibinfo {pages} {8} (\bibinfo {year} {2016})}\BibitemShut
  {NoStop}%
\bibitem [{\citenamefont {Lu}\ \emph {et~al.}(2023)\citenamefont {Lu},
  \citenamefont {Liscidini}, \citenamefont {Gaeta}, \citenamefont {Weiner},\
  and\ \citenamefont {Lukens}}]{2023_Lu_FB_review}%
  \BibitemOpen
  \bibfield  {author} {\bibinfo {author} {\bibfnamefont {H.-H.}\ \bibnamefont
  {Lu}}, \bibinfo {author} {\bibfnamefont {M.}~\bibnamefont {Liscidini}},
  \bibinfo {author} {\bibfnamefont {A.~L.}\ \bibnamefont {Gaeta}}, \bibinfo
  {author} {\bibfnamefont {A.~M.}\ \bibnamefont {Weiner}},\ and\ \bibinfo
  {author} {\bibfnamefont {J.~M.}\ \bibnamefont {Lukens}},\ }\bibfield  {title}
  {\bibinfo {title} {Frequency-bin photonic quantum information},\ }\href@noop
  {} {\bibfield  {journal} {\bibinfo  {journal} {Optica}\ }\textbf {\bibinfo
  {volume} {10}},\ \bibinfo {pages} {1655} (\bibinfo {year}
  {2023})}\BibitemShut {NoStop}%
\bibitem [{\citenamefont {Myilswamy}\ \emph {et~al.}(2025)\citenamefont
  {Myilswamy}, \citenamefont {Cohen}, \citenamefont {Seshadri}, \citenamefont
  {Lu},\ and\ \citenamefont {Lukens}}]{2025_Lukens_FBreview}%
  \BibitemOpen
  \bibfield  {author} {\bibinfo {author} {\bibfnamefont {K.~V.}\ \bibnamefont
  {Myilswamy}}, \bibinfo {author} {\bibfnamefont {L.~M.}\ \bibnamefont
  {Cohen}}, \bibinfo {author} {\bibfnamefont {S.}~\bibnamefont {Seshadri}},
  \bibinfo {author} {\bibfnamefont {H.-H.}\ \bibnamefont {Lu}},\ and\ \bibinfo
  {author} {\bibfnamefont {J.~M.}\ \bibnamefont {Lukens}},\ }\bibfield  {title}
  {\bibinfo {title} {On-chip frequency-bin quantum photonics},\ }\href@noop {}
  {\bibfield  {journal} {\bibinfo  {journal} {Nanophotonics}\ }\textbf
  {\bibinfo {volume} {14}},\ \bibinfo {pages} {1879} (\bibinfo {year}
  {2025})}\BibitemShut {NoStop}%
\bibitem [{\citenamefont {Wang}\ \emph {et~al.}(2025)\citenamefont {Wang},
  \citenamefont {Li}, \citenamefont {Wang}, \citenamefont {Zhou}, \citenamefont
  {Cheng}, \citenamefont {Jing}, \citenamefont {Sun}, \citenamefont {Li},
  \citenamefont {Li}, \citenamefont {Wu} \emph {et~al.}}]{wang2025large}%
  \BibitemOpen
  \bibfield  {author} {\bibinfo {author} {\bibfnamefont {Z.}~\bibnamefont
  {Wang}}, \bibinfo {author} {\bibfnamefont {K.}~\bibnamefont {Li}}, \bibinfo
  {author} {\bibfnamefont {Y.}~\bibnamefont {Wang}}, \bibinfo {author}
  {\bibfnamefont {X.}~\bibnamefont {Zhou}}, \bibinfo {author} {\bibfnamefont
  {Y.}~\bibnamefont {Cheng}}, \bibinfo {author} {\bibfnamefont
  {B.}~\bibnamefont {Jing}}, \bibinfo {author} {\bibfnamefont {F.}~\bibnamefont
  {Sun}}, \bibinfo {author} {\bibfnamefont {J.}~\bibnamefont {Li}}, \bibinfo
  {author} {\bibfnamefont {Z.}~\bibnamefont {Li}}, \bibinfo {author}
  {\bibfnamefont {B.}~\bibnamefont {Wu}}, \emph {et~al.},\ }\bibfield  {title}
  {\bibinfo {title} {Large-scale cluster quantum microcombs},\ }\href@noop {}
  {\bibfield  {journal} {\bibinfo  {journal} {Light: Science \& Applications}\
  }\textbf {\bibinfo {volume} {14}},\ \bibinfo {pages} {164} (\bibinfo {year}
  {2025})}\BibitemShut {NoStop}%
\bibitem [{\citenamefont {Imany}\ \emph {et~al.}(2019)\citenamefont {Imany},
  \citenamefont {Jaramillo-Villegas}, \citenamefont {Alshaykh}, \citenamefont
  {Lukens}, \citenamefont {Odele}, \citenamefont {Moore}, \citenamefont
  {Leaird}, \citenamefont {Qi},\ and\ \citenamefont {Weiner}}]{imany2019high}%
  \BibitemOpen
  \bibfield  {author} {\bibinfo {author} {\bibfnamefont {P.}~\bibnamefont
  {Imany}}, \bibinfo {author} {\bibfnamefont {J.~A.}\ \bibnamefont
  {Jaramillo-Villegas}}, \bibinfo {author} {\bibfnamefont {M.~S.}\ \bibnamefont
  {Alshaykh}}, \bibinfo {author} {\bibfnamefont {J.~M.}\ \bibnamefont
  {Lukens}}, \bibinfo {author} {\bibfnamefont {O.~D.}\ \bibnamefont {Odele}},
  \bibinfo {author} {\bibfnamefont {A.~J.}\ \bibnamefont {Moore}}, \bibinfo
  {author} {\bibfnamefont {D.~E.}\ \bibnamefont {Leaird}}, \bibinfo {author}
  {\bibfnamefont {M.}~\bibnamefont {Qi}},\ and\ \bibinfo {author}
  {\bibfnamefont {A.~M.}\ \bibnamefont {Weiner}},\ }\bibfield  {title}
  {\bibinfo {title} {High-dimensional optical quantum logic in large
  operational spaces},\ }\href@noop {} {\bibfield  {journal} {\bibinfo
  {journal} {npj Quantum Information}\ }\textbf {\bibinfo {volume} {5}},\
  \bibinfo {pages} {59} (\bibinfo {year} {2019})}\BibitemShut {NoStop}%
\bibitem [{\citenamefont {Henry}\ \emph {et~al.}(2023)\citenamefont {Henry},
  \citenamefont {Raghunathan}, \citenamefont {Ricard}, \citenamefont
  {Lefaucher}, \citenamefont {Miatto}, \citenamefont {Belabas}, \citenamefont
  {Zaquine},\ and\ \citenamefont {All{\'e}aume}}]{henry2023parallelizable}%
  \BibitemOpen
  \bibfield  {author} {\bibinfo {author} {\bibfnamefont {A.}~\bibnamefont
  {Henry}}, \bibinfo {author} {\bibfnamefont {R.}~\bibnamefont {Raghunathan}},
  \bibinfo {author} {\bibfnamefont {G.}~\bibnamefont {Ricard}}, \bibinfo
  {author} {\bibfnamefont {B.}~\bibnamefont {Lefaucher}}, \bibinfo {author}
  {\bibfnamefont {F.}~\bibnamefont {Miatto}}, \bibinfo {author} {\bibfnamefont
  {N.}~\bibnamefont {Belabas}}, \bibinfo {author} {\bibfnamefont
  {I.}~\bibnamefont {Zaquine}},\ and\ \bibinfo {author} {\bibfnamefont
  {R.}~\bibnamefont {All{\'e}aume}},\ }\bibfield  {title} {\bibinfo {title}
  {Parallelizable synthesis of arbitrary single-qubit gates with linear optics
  and time-frequency encoding},\ }\href@noop {} {\bibfield  {journal} {\bibinfo
   {journal} {Physical Review A}\ }\textbf {\bibinfo {volume} {107}},\ \bibinfo
  {pages} {062610} (\bibinfo {year} {2023})}\BibitemShut {NoStop}%
\bibitem [{\citenamefont {Henry}\ \emph {et~al.}(2024)\citenamefont {Henry},
  \citenamefont {Fioretto}, \citenamefont {Procopio}, \citenamefont {Monfray},
  \citenamefont {Boeuf}, \citenamefont {Vivien}, \citenamefont {Cassan},
  \citenamefont {Alonzo-Ramos}, \citenamefont {Bencheikh}, \citenamefont
  {Zaquine} \emph {et~al.}}]{henry2024parallelization}%
  \BibitemOpen
  \bibfield  {author} {\bibinfo {author} {\bibfnamefont {A.}~\bibnamefont
  {Henry}}, \bibinfo {author} {\bibfnamefont {D.~A.}\ \bibnamefont {Fioretto}},
  \bibinfo {author} {\bibfnamefont {L.~M.}\ \bibnamefont {Procopio}}, \bibinfo
  {author} {\bibfnamefont {S.}~\bibnamefont {Monfray}}, \bibinfo {author}
  {\bibfnamefont {F.}~\bibnamefont {Boeuf}}, \bibinfo {author} {\bibfnamefont
  {L.}~\bibnamefont {Vivien}}, \bibinfo {author} {\bibfnamefont
  {E.}~\bibnamefont {Cassan}}, \bibinfo {author} {\bibfnamefont
  {C.}~\bibnamefont {Alonzo-Ramos}}, \bibinfo {author} {\bibfnamefont
  {K.}~\bibnamefont {Bencheikh}}, \bibinfo {author} {\bibfnamefont
  {I.}~\bibnamefont {Zaquine}}, \emph {et~al.},\ }\bibfield  {title} {\bibinfo
  {title} {Parallelization of frequency domain quantum gates: manipulation and
  distribution of frequency-entangled photon pairs generated by a 21 ghz
  silicon microresonator},\ }\href@noop {} {\bibfield  {journal} {\bibinfo
  {journal} {Advanced Photonics}\ }\textbf {\bibinfo {volume} {6}},\ \bibinfo
  {pages} {036003} (\bibinfo {year} {2024})}\BibitemShut {NoStop}%
\bibitem [{\citenamefont {Chapman}\ \emph {et~al.}(2025)\citenamefont
  {Chapman}, \citenamefont {Seshadri}, \citenamefont {Lukens}, \citenamefont
  {Peters}, \citenamefont {McKinney}, \citenamefont {Weiner},\ and\
  \citenamefont {Lu}}]{chapman2025quantum}%
  \BibitemOpen
  \bibfield  {author} {\bibinfo {author} {\bibfnamefont {S.~D.}\ \bibnamefont
  {Chapman}}, \bibinfo {author} {\bibfnamefont {S.}~\bibnamefont {Seshadri}},
  \bibinfo {author} {\bibfnamefont {J.~M.}\ \bibnamefont {Lukens}}, \bibinfo
  {author} {\bibfnamefont {N.~A.}\ \bibnamefont {Peters}}, \bibinfo {author}
  {\bibfnamefont {J.~D.}\ \bibnamefont {McKinney}}, \bibinfo {author}
  {\bibfnamefont {A.~M.}\ \bibnamefont {Weiner}},\ and\ \bibinfo {author}
  {\bibfnamefont {H.-H.}\ \bibnamefont {Lu}},\ }\bibfield  {title} {\bibinfo
  {title} {Quantum nonlocal modulation cancelation with distributed clocks},\
  }\href@noop {} {\bibfield  {journal} {\bibinfo  {journal} {Optica Quantum}\
  }\textbf {\bibinfo {volume} {3}},\ \bibinfo {pages} {45} (\bibinfo {year}
  {2025})}\BibitemShut {NoStop}%
\bibitem [{\citenamefont {Kues}\ \emph {et~al.}(2017)\citenamefont {Kues},
  \citenamefont {Reimer}, \citenamefont {Roztocki}, \citenamefont {Cort{\'e}s},
  \citenamefont {Sciara}, \citenamefont {Wetzel}, \citenamefont {Zhang},
  \citenamefont {Cino}, \citenamefont {Chu}, \citenamefont {Little} \emph
  {et~al.}}]{2017_Kues_FB}%
  \BibitemOpen
  \bibfield  {author} {\bibinfo {author} {\bibfnamefont {M.}~\bibnamefont
  {Kues}}, \bibinfo {author} {\bibfnamefont {C.}~\bibnamefont {Reimer}},
  \bibinfo {author} {\bibfnamefont {P.}~\bibnamefont {Roztocki}}, \bibinfo
  {author} {\bibfnamefont {L.~R.}\ \bibnamefont {Cort{\'e}s}}, \bibinfo
  {author} {\bibfnamefont {S.}~\bibnamefont {Sciara}}, \bibinfo {author}
  {\bibfnamefont {B.}~\bibnamefont {Wetzel}}, \bibinfo {author} {\bibfnamefont
  {Y.}~\bibnamefont {Zhang}}, \bibinfo {author} {\bibfnamefont
  {A.}~\bibnamefont {Cino}}, \bibinfo {author} {\bibfnamefont {S.~T.}\
  \bibnamefont {Chu}}, \bibinfo {author} {\bibfnamefont {B.~E.}\ \bibnamefont
  {Little}}, \emph {et~al.},\ }\bibfield  {title} {\bibinfo {title} {On-chip
  generation of high-dimensional entangled quantum states and their coherent
  control},\ }\href@noop {} {\bibfield  {journal} {\bibinfo  {journal}
  {Nature}\ }\textbf {\bibinfo {volume} {546}},\ \bibinfo {pages} {622}
  (\bibinfo {year} {2017})}\BibitemShut {NoStop}%
\bibitem [{\citenamefont {Fabre}\ \emph {et~al.}(2022)\citenamefont {Fabre},
  \citenamefont {Keller},\ and\ \citenamefont {Milman}}]{fabre2022time}%
  \BibitemOpen
  \bibfield  {author} {\bibinfo {author} {\bibfnamefont {N.}~\bibnamefont
  {Fabre}}, \bibinfo {author} {\bibfnamefont {A.}~\bibnamefont {Keller}},\ and\
  \bibinfo {author} {\bibfnamefont {P.}~\bibnamefont {Milman}},\ }\bibfield
  {title} {\bibinfo {title} {Time and frequency as quantum continuous
  variables},\ }\href@noop {} {\bibfield  {journal} {\bibinfo  {journal}
  {Physical Review A}\ }\textbf {\bibinfo {volume} {105}},\ \bibinfo {pages}
  {052429} (\bibinfo {year} {2022})}\BibitemShut {NoStop}%
\bibitem [{\citenamefont {Lukens}\ \emph {et~al.}(2026)\citenamefont {Lukens},
  \citenamefont {Dallyn}, \citenamefont {Lu}, \citenamefont {Wasserbeck},
  \citenamefont {Graf}, \citenamefont {Gehl}, \citenamefont {Davids},\ and\
  \citenamefont {Otterstrom}}]{lukens2026paradigm}%
  \BibitemOpen
  \bibfield  {author} {\bibinfo {author} {\bibfnamefont {J.~M.}\ \bibnamefont
  {Lukens}}, \bibinfo {author} {\bibfnamefont {J.~H.}\ \bibnamefont {Dallyn}},
  \bibinfo {author} {\bibfnamefont {H.-H.}\ \bibnamefont {Lu}}, \bibinfo
  {author} {\bibfnamefont {N.~I.}\ \bibnamefont {Wasserbeck}}, \bibinfo
  {author} {\bibfnamefont {A.~J.}\ \bibnamefont {Graf}}, \bibinfo {author}
  {\bibfnamefont {M.}~\bibnamefont {Gehl}}, \bibinfo {author} {\bibfnamefont
  {P.~S.}\ \bibnamefont {Davids}},\ and\ \bibinfo {author} {\bibfnamefont
  {N.~T.}\ \bibnamefont {Otterstrom}},\ }\bibfield  {title} {\bibinfo {title}
  {A paradigm for universal quantum information processing with integrated
  acousto-optic frequency beamsplitters},\ }\href@noop {} {\bibfield  {journal}
  {\bibinfo  {journal} {arXiv preprint arXiv:2601.06752}\ } (\bibinfo {year}
  {2026})}\BibitemShut {NoStop}%
\bibitem [{\citenamefont {Borghi}\ \emph {et~al.}(2023)\citenamefont {Borghi},
  \citenamefont {Tagliavacche}, \citenamefont {Sabattoli}, \citenamefont
  {Dirani}, \citenamefont {Youssef}, \citenamefont {Petit-Etienne},
  \citenamefont {Pargon}, \citenamefont {Sipe}, \citenamefont {Liscidini},
  \citenamefont {Sciancalepore} \emph {et~al.}}]{2023_Borghi}%
  \BibitemOpen
  \bibfield  {author} {\bibinfo {author} {\bibfnamefont {M.}~\bibnamefont
  {Borghi}}, \bibinfo {author} {\bibfnamefont {N.}~\bibnamefont
  {Tagliavacche}}, \bibinfo {author} {\bibfnamefont {F.~A.}\ \bibnamefont
  {Sabattoli}}, \bibinfo {author} {\bibfnamefont {H.~E.}\ \bibnamefont
  {Dirani}}, \bibinfo {author} {\bibfnamefont {L.}~\bibnamefont {Youssef}},
  \bibinfo {author} {\bibfnamefont {C.}~\bibnamefont {Petit-Etienne}}, \bibinfo
  {author} {\bibfnamefont {E.}~\bibnamefont {Pargon}}, \bibinfo {author}
  {\bibfnamefont {J.}~\bibnamefont {Sipe}}, \bibinfo {author} {\bibfnamefont
  {M.}~\bibnamefont {Liscidini}}, \bibinfo {author} {\bibfnamefont
  {C.}~\bibnamefont {Sciancalepore}}, \emph {et~al.},\ }\bibfield  {title}
  {\bibinfo {title} {Reconfigurable silicon photonic chip for the generation of
  frequency-bin-entangled qudits},\ }\href@noop {} {\bibfield  {journal}
  {\bibinfo  {journal} {Physical Review Applied}\ }\textbf {\bibinfo {volume}
  {19}},\ \bibinfo {pages} {064026} (\bibinfo {year} {2023})}\BibitemShut
  {NoStop}%
\bibitem [{\citenamefont {Clementi}\ \emph {et~al.}(2023)\citenamefont
  {Clementi}, \citenamefont {Sabattoli}, \citenamefont {Borghi}, \citenamefont
  {Gianini}, \citenamefont {Tagliavacche}, \citenamefont {El~Dirani},
  \citenamefont {Youssef}, \citenamefont {Bergamasco}, \citenamefont
  {Petit-Etienne}, \citenamefont {Pargon} \emph
  {et~al.}}]{clementi2023programmable}%
  \BibitemOpen
  \bibfield  {author} {\bibinfo {author} {\bibfnamefont {M.}~\bibnamefont
  {Clementi}}, \bibinfo {author} {\bibfnamefont {F.~A.}\ \bibnamefont
  {Sabattoli}}, \bibinfo {author} {\bibfnamefont {M.}~\bibnamefont {Borghi}},
  \bibinfo {author} {\bibfnamefont {L.}~\bibnamefont {Gianini}}, \bibinfo
  {author} {\bibfnamefont {N.}~\bibnamefont {Tagliavacche}}, \bibinfo {author}
  {\bibfnamefont {H.}~\bibnamefont {El~Dirani}}, \bibinfo {author}
  {\bibfnamefont {L.}~\bibnamefont {Youssef}}, \bibinfo {author} {\bibfnamefont
  {N.}~\bibnamefont {Bergamasco}}, \bibinfo {author} {\bibfnamefont
  {C.}~\bibnamefont {Petit-Etienne}}, \bibinfo {author} {\bibfnamefont
  {E.}~\bibnamefont {Pargon}}, \emph {et~al.},\ }\bibfield  {title} {\bibinfo
  {title} {Programmable frequency-bin quantum states in a nano-engineered
  silicon device},\ }\href@noop {} {\bibfield  {journal} {\bibinfo  {journal}
  {Nature Communications}\ }\textbf {\bibinfo {volume} {14}},\ \bibinfo {pages}
  {176} (\bibinfo {year} {2023})}\BibitemShut {NoStop}%
\bibitem [{\citenamefont {Lu}\ \emph {et~al.}(2022{\natexlab{a}})\citenamefont
  {Lu}, \citenamefont {Myilswamy}, \citenamefont {Bennink}, \citenamefont
  {Seshadri}, \citenamefont {Alshaykh}, \citenamefont {Liu}, \citenamefont
  {Kippenberg}, \citenamefont {Leaird}, \citenamefont {Weiner},\ and\
  \citenamefont {Lukens}}]{lu2022bayesian}%
  \BibitemOpen
  \bibfield  {author} {\bibinfo {author} {\bibfnamefont {H.-H.}\ \bibnamefont
  {Lu}}, \bibinfo {author} {\bibfnamefont {K.~V.}\ \bibnamefont {Myilswamy}},
  \bibinfo {author} {\bibfnamefont {R.~S.}\ \bibnamefont {Bennink}}, \bibinfo
  {author} {\bibfnamefont {S.}~\bibnamefont {Seshadri}}, \bibinfo {author}
  {\bibfnamefont {M.~S.}\ \bibnamefont {Alshaykh}}, \bibinfo {author}
  {\bibfnamefont {J.}~\bibnamefont {Liu}}, \bibinfo {author} {\bibfnamefont
  {T.~J.}\ \bibnamefont {Kippenberg}}, \bibinfo {author} {\bibfnamefont
  {D.~E.}\ \bibnamefont {Leaird}}, \bibinfo {author} {\bibfnamefont {A.~M.}\
  \bibnamefont {Weiner}},\ and\ \bibinfo {author} {\bibfnamefont {J.~M.}\
  \bibnamefont {Lukens}},\ }\bibfield  {title} {\bibinfo {title} {Bayesian
  tomography of high-dimensional on-chip biphoton frequency combs with
  randomized measurements},\ }\href@noop {} {\bibfield  {journal} {\bibinfo
  {journal} {Nature Communications}\ }\textbf {\bibinfo {volume} {13}},\
  \bibinfo {pages} {4338} (\bibinfo {year} {2022}{\natexlab{a}})}\BibitemShut
  {NoStop}%
\bibitem [{\citenamefont {Joshi}\ \emph {et~al.}(2020)\citenamefont {Joshi},
  \citenamefont {Farsi}, \citenamefont {Dutt}, \citenamefont {Kim},
  \citenamefont {Ji}, \citenamefont {Zhao}, \citenamefont {Bishop},
  \citenamefont {Lipson},\ and\ \citenamefont {Gaeta}}]{joshi2020frequency}%
  \BibitemOpen
  \bibfield  {author} {\bibinfo {author} {\bibfnamefont {C.}~\bibnamefont
  {Joshi}}, \bibinfo {author} {\bibfnamefont {A.}~\bibnamefont {Farsi}},
  \bibinfo {author} {\bibfnamefont {A.}~\bibnamefont {Dutt}}, \bibinfo {author}
  {\bibfnamefont {B.~Y.}\ \bibnamefont {Kim}}, \bibinfo {author} {\bibfnamefont
  {X.}~\bibnamefont {Ji}}, \bibinfo {author} {\bibfnamefont {Y.}~\bibnamefont
  {Zhao}}, \bibinfo {author} {\bibfnamefont {A.~M.}\ \bibnamefont {Bishop}},
  \bibinfo {author} {\bibfnamefont {M.}~\bibnamefont {Lipson}},\ and\ \bibinfo
  {author} {\bibfnamefont {A.~L.}\ \bibnamefont {Gaeta}},\ }\bibfield  {title}
  {\bibinfo {title} {Frequency-domain quantum interference with correlated
  photons from an integrated microresonator},\ }\href@noop {} {\bibfield
  {journal} {\bibinfo  {journal} {Physical review letters}\ }\textbf {\bibinfo
  {volume} {124}},\ \bibinfo {pages} {143601} (\bibinfo {year}
  {2020})}\BibitemShut {NoStop}%
\bibitem [{\citenamefont {Jing}\ \emph {et~al.}(2025)\citenamefont {Jing},
  \citenamefont {Fan}, \citenamefont {Zheng}, \citenamefont {Chen},
  \citenamefont {Kong}, \citenamefont {Niu},\ and\ \citenamefont
  {Lu}}]{jing2025hong}%
  \BibitemOpen
  \bibfield  {author} {\bibinfo {author} {\bibfnamefont {X.}~\bibnamefont
  {Jing}}, \bibinfo {author} {\bibfnamefont {L.}~\bibnamefont {Fan}}, \bibinfo
  {author} {\bibfnamefont {X.}~\bibnamefont {Zheng}}, \bibinfo {author}
  {\bibfnamefont {T.}~\bibnamefont {Chen}}, \bibinfo {author} {\bibfnamefont
  {Y.}~\bibnamefont {Kong}}, \bibinfo {author} {\bibfnamefont {B.}~\bibnamefont
  {Niu}},\ and\ \bibinfo {author} {\bibfnamefont {L.}~\bibnamefont {Lu}},\
  }\bibfield  {title} {\bibinfo {title} {Hong--ou--mandel interferometry and
  quantum metrology with multimode frequency-bin entangled photons},\
  }\href@noop {} {\bibfield  {journal} {\bibinfo  {journal} {APL Photonics}\
  }\textbf {\bibinfo {volume} {10}} (\bibinfo {year} {2025})}\BibitemShut
  {NoStop}%
\bibitem [{\citenamefont {Seshadri}\ \emph {et~al.}(2022)\citenamefont
  {Seshadri}, \citenamefont {Lingaraju}, \citenamefont {Lu}, \citenamefont
  {Imany}, \citenamefont {Leaird},\ and\ \citenamefont
  {Weiner}}]{seshadri2022nonlocal}%
  \BibitemOpen
  \bibfield  {author} {\bibinfo {author} {\bibfnamefont {S.}~\bibnamefont
  {Seshadri}}, \bibinfo {author} {\bibfnamefont {N.}~\bibnamefont {Lingaraju}},
  \bibinfo {author} {\bibfnamefont {H.-H.}\ \bibnamefont {Lu}}, \bibinfo
  {author} {\bibfnamefont {P.}~\bibnamefont {Imany}}, \bibinfo {author}
  {\bibfnamefont {D.~E.}\ \bibnamefont {Leaird}},\ and\ \bibinfo {author}
  {\bibfnamefont {A.~M.}\ \bibnamefont {Weiner}},\ }\bibfield  {title}
  {\bibinfo {title} {Nonlocal subpicosecond delay metrology using spectral
  quantum interference},\ }\href@noop {} {\bibfield  {journal} {\bibinfo
  {journal} {Optica}\ }\textbf {\bibinfo {volume} {9}},\ \bibinfo {pages}
  {1339} (\bibinfo {year} {2022})}\BibitemShut {NoStop}%
\bibitem [{\citenamefont {Imany}\ \emph {et~al.}(2020)\citenamefont {Imany},
  \citenamefont {Lingaraju}, \citenamefont {Alshaykh}, \citenamefont {Leaird},\
  and\ \citenamefont {Weiner}}]{2020_Imany_QWalk}%
  \BibitemOpen
  \bibfield  {author} {\bibinfo {author} {\bibfnamefont {P.}~\bibnamefont
  {Imany}}, \bibinfo {author} {\bibfnamefont {N.~B.}\ \bibnamefont
  {Lingaraju}}, \bibinfo {author} {\bibfnamefont {M.~S.}\ \bibnamefont
  {Alshaykh}}, \bibinfo {author} {\bibfnamefont {D.~E.}\ \bibnamefont
  {Leaird}},\ and\ \bibinfo {author} {\bibfnamefont {A.~M.}\ \bibnamefont
  {Weiner}},\ }\bibfield  {title} {\bibinfo {title} {Probing quantum walks
  through coherent control of high-dimensionally entangled photons},\ }\href
  {https://doi.org/10.1126/sciadv.aba8066} {\bibfield  {journal} {\bibinfo
  {journal} {Science Advances}\ }\textbf {\bibinfo {volume} {6}},\ \bibinfo
  {pages} {eaba8066} (\bibinfo {year} {2020})}\BibitemShut {NoStop}%
\bibitem [{\citenamefont {Haldar}\ \emph {et~al.}(2022)\citenamefont {Haldar},
  \citenamefont {Johanning}, \citenamefont {R{\"u}beling}, \citenamefont
  {Khodadad~Kashi}, \citenamefont {B{\ae}kkegaard}, \citenamefont {Bose},
  \citenamefont {Zinner},\ and\ \citenamefont {Kues}}]{2022_Haldar_QWalk}%
  \BibitemOpen
  \bibfield  {author} {\bibinfo {author} {\bibfnamefont {R.}~\bibnamefont
  {Haldar}}, \bibinfo {author} {\bibfnamefont {R.}~\bibnamefont {Johanning}},
  \bibinfo {author} {\bibfnamefont {P.}~\bibnamefont {R{\"u}beling}}, \bibinfo
  {author} {\bibfnamefont {A.}~\bibnamefont {Khodadad~Kashi}}, \bibinfo
  {author} {\bibfnamefont {T.}~\bibnamefont {B{\ae}kkegaard}}, \bibinfo
  {author} {\bibfnamefont {S.}~\bibnamefont {Bose}}, \bibinfo {author}
  {\bibfnamefont {N.~T.}\ \bibnamefont {Zinner}},\ and\ \bibinfo {author}
  {\bibfnamefont {M.}~\bibnamefont {Kues}},\ }\bibfield  {title} {\bibinfo
  {title} {Steering of quantum walks through coherent control of
  high-dimensional bi-photon quantum frequency combs with tunable state
  entropies},\ }\href@noop {} {\bibfield  {journal} {\bibinfo  {journal} {arXiv
  preprint arXiv:2210.06305}\ } (\bibinfo {year} {2022})}\BibitemShut {NoStop}%
\bibitem [{\citenamefont {Khodadad~Kashi}\ and\ \citenamefont
  {Kues}(2025)}]{2025_Kues_FB_QKD}%
  \BibitemOpen
  \bibfield  {author} {\bibinfo {author} {\bibfnamefont {A.}~\bibnamefont
  {Khodadad~Kashi}}\ and\ \bibinfo {author} {\bibfnamefont {M.}~\bibnamefont
  {Kues}},\ }\bibfield  {title} {\bibinfo {title} {Frequency-bin-encoded
  entanglement-based quantum key distribution in a reconfigurable
  frequency-multiplexed network},\ }\href@noop {} {\bibfield  {journal}
  {\bibinfo  {journal} {Light: Science \& Applications}\ }\textbf {\bibinfo
  {volume} {14}},\ \bibinfo {pages} {49} (\bibinfo {year} {2025})}\BibitemShut
  {NoStop}%
\bibitem [{\citenamefont {Crisan}\ \emph {et~al.}(2025)\citenamefont {Crisan},
  \citenamefont {Henry}, \citenamefont {Fioretto}, \citenamefont {Alvarez},
  \citenamefont {Monfray}, \citenamefont {Boeuf}, \citenamefont {Vivien},
  \citenamefont {Cassan}, \citenamefont {Alonso-Ramos},\ and\ \citenamefont
  {Belabas}}]{2025_Belabas_FB_QKD}%
  \BibitemOpen
  \bibfield  {author} {\bibinfo {author} {\bibfnamefont {G.~C.}\ \bibnamefont
  {Crisan}}, \bibinfo {author} {\bibfnamefont {A.}~\bibnamefont {Henry}},
  \bibinfo {author} {\bibfnamefont {D.~A.}\ \bibnamefont {Fioretto}}, \bibinfo
  {author} {\bibfnamefont {J.~R.}\ \bibnamefont {Alvarez}}, \bibinfo {author}
  {\bibfnamefont {S.}~\bibnamefont {Monfray}}, \bibinfo {author} {\bibfnamefont
  {F.}~\bibnamefont {Boeuf}}, \bibinfo {author} {\bibfnamefont
  {L.}~\bibnamefont {Vivien}}, \bibinfo {author} {\bibfnamefont
  {E.}~\bibnamefont {Cassan}}, \bibinfo {author} {\bibfnamefont
  {C.}~\bibnamefont {Alonso-Ramos}},\ and\ \bibinfo {author} {\bibfnamefont
  {N.}~\bibnamefont {Belabas}},\ }\bibfield  {title} {\bibinfo {title}
  {Multi-dimensional frequency-bin entanglement-based quantum key distribution
  network},\ }\href@noop {} {\bibfield  {journal} {\bibinfo  {journal} {arXiv
  preprint arXiv:2507.00972}\ } (\bibinfo {year} {2025})}\BibitemShut {NoStop}%
\bibitem [{\citenamefont {Tagliavacche}\ \emph {et~al.}(2025)\citenamefont
  {Tagliavacche}, \citenamefont {Borghi}, \citenamefont {Guarda}, \citenamefont
  {Ribezzo}, \citenamefont {Liscidini}, \citenamefont {Bacco}, \citenamefont
  {Galli},\ and\ \citenamefont {Bajoni}}]{2024_Tagliavacche}%
  \BibitemOpen
  \bibfield  {author} {\bibinfo {author} {\bibfnamefont {N.}~\bibnamefont
  {Tagliavacche}}, \bibinfo {author} {\bibfnamefont {M.}~\bibnamefont
  {Borghi}}, \bibinfo {author} {\bibfnamefont {G.}~\bibnamefont {Guarda}},
  \bibinfo {author} {\bibfnamefont {D.}~\bibnamefont {Ribezzo}}, \bibinfo
  {author} {\bibfnamefont {M.}~\bibnamefont {Liscidini}}, \bibinfo {author}
  {\bibfnamefont {D.}~\bibnamefont {Bacco}}, \bibinfo {author} {\bibfnamefont
  {M.}~\bibnamefont {Galli}},\ and\ \bibinfo {author} {\bibfnamefont
  {D.}~\bibnamefont {Bajoni}},\ }\bibfield  {title} {\bibinfo {title}
  {Frequency-bin entanglement-based quantum key distribution},\ }\href@noop {}
  {\bibfield  {journal} {\bibinfo  {journal} {npj Quantum Information}\
  }\textbf {\bibinfo {volume} {11}},\ \bibinfo {pages} {60} (\bibinfo {year}
  {2025})}\BibitemShut {NoStop}%
\bibitem [{\citenamefont {Jia}\ \emph {et~al.}(2025)\citenamefont {Jia},
  \citenamefont {Zhai}, \citenamefont {Zhu}, \citenamefont {You}, \citenamefont
  {Cao}, \citenamefont {Zhang}, \citenamefont {Zheng}, \citenamefont {Fu},
  \citenamefont {Mao}, \citenamefont {Dai} \emph {et~al.}}]{jia2025continuous}%
  \BibitemOpen
  \bibfield  {author} {\bibinfo {author} {\bibfnamefont {X.}~\bibnamefont
  {Jia}}, \bibinfo {author} {\bibfnamefont {C.}~\bibnamefont {Zhai}}, \bibinfo
  {author} {\bibfnamefont {X.}~\bibnamefont {Zhu}}, \bibinfo {author}
  {\bibfnamefont {C.}~\bibnamefont {You}}, \bibinfo {author} {\bibfnamefont
  {Y.}~\bibnamefont {Cao}}, \bibinfo {author} {\bibfnamefont {X.}~\bibnamefont
  {Zhang}}, \bibinfo {author} {\bibfnamefont {Y.}~\bibnamefont {Zheng}},
  \bibinfo {author} {\bibfnamefont {Z.}~\bibnamefont {Fu}}, \bibinfo {author}
  {\bibfnamefont {J.}~\bibnamefont {Mao}}, \bibinfo {author} {\bibfnamefont
  {T.}~\bibnamefont {Dai}}, \emph {et~al.},\ }\bibfield  {title} {\bibinfo
  {title} {Continuous-variable multipartite entanglement in an integrated
  microcomb},\ }\href@noop {} {\bibfield  {journal} {\bibinfo  {journal}
  {Nature}\ ,\ \bibinfo {pages} {1}} (\bibinfo {year} {2025})}\BibitemShut
  {NoStop}%
\bibitem [{\citenamefont {Lu}\ \emph {et~al.}(2018{\natexlab{a}})\citenamefont
  {Lu}, \citenamefont {Lukens}, \citenamefont {Peters}, \citenamefont {Odele},
  \citenamefont {Leaird}, \citenamefont {Weiner},\ and\ \citenamefont
  {Lougovski}}]{2018_Lu_FB_BS}%
  \BibitemOpen
  \bibfield  {author} {\bibinfo {author} {\bibfnamefont {H.-H.}\ \bibnamefont
  {Lu}}, \bibinfo {author} {\bibfnamefont {J.~M.}\ \bibnamefont {Lukens}},
  \bibinfo {author} {\bibfnamefont {N.~A.}\ \bibnamefont {Peters}}, \bibinfo
  {author} {\bibfnamefont {O.~D.}\ \bibnamefont {Odele}}, \bibinfo {author}
  {\bibfnamefont {D.~E.}\ \bibnamefont {Leaird}}, \bibinfo {author}
  {\bibfnamefont {A.~M.}\ \bibnamefont {Weiner}},\ and\ \bibinfo {author}
  {\bibfnamefont {P.}~\bibnamefont {Lougovski}},\ }\bibfield  {title} {\bibinfo
  {title} {Electro-optic frequency beam splitters and tritters for
  high-fidelity photonic quantum information processing},\ }\href@noop {}
  {\bibfield  {journal} {\bibinfo  {journal} {Physical review letters}\
  }\textbf {\bibinfo {volume} {120}},\ \bibinfo {pages} {030502} (\bibinfo
  {year} {2018}{\natexlab{a}})}\BibitemShut {NoStop}%
\bibitem [{\citenamefont {Lu}\ \emph {et~al.}(2022{\natexlab{b}})\citenamefont
  {Lu}, \citenamefont {Lingaraju}, \citenamefont {Leaird}, \citenamefont
  {Weiner},\ and\ \citenamefont {Lukens}}]{lu2022high}%
  \BibitemOpen
  \bibfield  {author} {\bibinfo {author} {\bibfnamefont {H.-H.}\ \bibnamefont
  {Lu}}, \bibinfo {author} {\bibfnamefont {N.~B.}\ \bibnamefont {Lingaraju}},
  \bibinfo {author} {\bibfnamefont {D.~E.}\ \bibnamefont {Leaird}}, \bibinfo
  {author} {\bibfnamefont {A.~M.}\ \bibnamefont {Weiner}},\ and\ \bibinfo
  {author} {\bibfnamefont {J.~M.}\ \bibnamefont {Lukens}},\ }\bibfield  {title}
  {\bibinfo {title} {High-dimensional discrete fourier transform gates with a
  quantum frequency processor},\ }\href@noop {} {\bibfield  {journal} {\bibinfo
   {journal} {Optics Express}\ }\textbf {\bibinfo {volume} {30}},\ \bibinfo
  {pages} {10126} (\bibinfo {year} {2022}{\natexlab{b}})}\BibitemShut {NoStop}%
\bibitem [{\citenamefont {Lingaraju}\ \emph {et~al.}(2022)\citenamefont
  {Lingaraju}, \citenamefont {Lu}, \citenamefont {Leaird}, \citenamefont
  {Estrella}, \citenamefont {Lukens},\ and\ \citenamefont
  {Weiner}}]{lingaraju2022bell}%
  \BibitemOpen
  \bibfield  {author} {\bibinfo {author} {\bibfnamefont {N.~B.}\ \bibnamefont
  {Lingaraju}}, \bibinfo {author} {\bibfnamefont {H.-H.}\ \bibnamefont {Lu}},
  \bibinfo {author} {\bibfnamefont {D.~E.}\ \bibnamefont {Leaird}}, \bibinfo
  {author} {\bibfnamefont {S.}~\bibnamefont {Estrella}}, \bibinfo {author}
  {\bibfnamefont {J.~M.}\ \bibnamefont {Lukens}},\ and\ \bibinfo {author}
  {\bibfnamefont {A.~M.}\ \bibnamefont {Weiner}},\ }\bibfield  {title}
  {\bibinfo {title} {Bell state analyzer for spectrally distinct photons},\
  }\href@noop {} {\bibfield  {journal} {\bibinfo  {journal} {Optica}\ }\textbf
  {\bibinfo {volume} {9}},\ \bibinfo {pages} {280} (\bibinfo {year}
  {2022})}\BibitemShut {NoStop}%
\bibitem [{\citenamefont {Lu}\ \emph {et~al.}(2019)\citenamefont {Lu},
  \citenamefont {Lukens}, \citenamefont {Williams}, \citenamefont {Imany},
  \citenamefont {Peters}, \citenamefont {Weiner},\ and\ \citenamefont
  {Lougovski}}]{lu2019controlled}%
  \BibitemOpen
  \bibfield  {author} {\bibinfo {author} {\bibfnamefont {H.-H.}\ \bibnamefont
  {Lu}}, \bibinfo {author} {\bibfnamefont {J.~M.}\ \bibnamefont {Lukens}},
  \bibinfo {author} {\bibfnamefont {B.~P.}\ \bibnamefont {Williams}}, \bibinfo
  {author} {\bibfnamefont {P.}~\bibnamefont {Imany}}, \bibinfo {author}
  {\bibfnamefont {N.~A.}\ \bibnamefont {Peters}}, \bibinfo {author}
  {\bibfnamefont {A.~M.}\ \bibnamefont {Weiner}},\ and\ \bibinfo {author}
  {\bibfnamefont {P.}~\bibnamefont {Lougovski}},\ }\bibfield  {title} {\bibinfo
  {title} {A controlled-not gate for frequency-bin qubits},\ }\href@noop {}
  {\bibfield  {journal} {\bibinfo  {journal} {npj Quantum Information}\
  }\textbf {\bibinfo {volume} {5}},\ \bibinfo {pages} {24} (\bibinfo {year}
  {2019})}\BibitemShut {NoStop}%
\bibitem [{\citenamefont {Zhu}\ \emph {et~al.}(2022)\citenamefont {Zhu},
  \citenamefont {Chen}, \citenamefont {Yu}, \citenamefont {Shao}, \citenamefont
  {Hu}, \citenamefont {Xin}, \citenamefont {Yeh}, \citenamefont {Ghosh},
  \citenamefont {He}, \citenamefont {Reimer} \emph {et~al.}}]{zhu2022spectral}%
  \BibitemOpen
  \bibfield  {author} {\bibinfo {author} {\bibfnamefont {D.}~\bibnamefont
  {Zhu}}, \bibinfo {author} {\bibfnamefont {C.}~\bibnamefont {Chen}}, \bibinfo
  {author} {\bibfnamefont {M.}~\bibnamefont {Yu}}, \bibinfo {author}
  {\bibfnamefont {L.}~\bibnamefont {Shao}}, \bibinfo {author} {\bibfnamefont
  {Y.}~\bibnamefont {Hu}}, \bibinfo {author} {\bibfnamefont {C.}~\bibnamefont
  {Xin}}, \bibinfo {author} {\bibfnamefont {M.}~\bibnamefont {Yeh}}, \bibinfo
  {author} {\bibfnamefont {S.}~\bibnamefont {Ghosh}}, \bibinfo {author}
  {\bibfnamefont {L.}~\bibnamefont {He}}, \bibinfo {author} {\bibfnamefont
  {C.}~\bibnamefont {Reimer}}, \emph {et~al.},\ }\bibfield  {title} {\bibinfo
  {title} {Spectral control of nonclassical light pulses using an integrated
  thin-film lithium niobate modulator},\ }\href@noop {} {\bibfield  {journal}
  {\bibinfo  {journal} {Light: Science \& Applications}\ }\textbf {\bibinfo
  {volume} {11}},\ \bibinfo {pages} {327} (\bibinfo {year} {2022})}\BibitemShut
  {NoStop}%
\bibitem [{\citenamefont {Assumpcao}\ \emph {et~al.}(2024)\citenamefont
  {Assumpcao}, \citenamefont {Renaud}, \citenamefont {Baradari}, \citenamefont
  {Zeng}, \citenamefont {De-Eknamkul}, \citenamefont {Xin}, \citenamefont
  {Shams-Ansari}, \citenamefont {Barton}, \citenamefont {Machielse},\ and\
  \citenamefont {Loncar}}]{assumpcao2024thin}%
  \BibitemOpen
  \bibfield  {author} {\bibinfo {author} {\bibfnamefont {D.}~\bibnamefont
  {Assumpcao}}, \bibinfo {author} {\bibfnamefont {D.}~\bibnamefont {Renaud}},
  \bibinfo {author} {\bibfnamefont {A.}~\bibnamefont {Baradari}}, \bibinfo
  {author} {\bibfnamefont {B.}~\bibnamefont {Zeng}}, \bibinfo {author}
  {\bibfnamefont {C.}~\bibnamefont {De-Eknamkul}}, \bibinfo {author}
  {\bibfnamefont {C.}~\bibnamefont {Xin}}, \bibinfo {author} {\bibfnamefont
  {A.}~\bibnamefont {Shams-Ansari}}, \bibinfo {author} {\bibfnamefont
  {D.}~\bibnamefont {Barton}}, \bibinfo {author} {\bibfnamefont
  {B.}~\bibnamefont {Machielse}},\ and\ \bibinfo {author} {\bibfnamefont
  {M.}~\bibnamefont {Loncar}},\ }\bibfield  {title} {\bibinfo {title} {A thin
  film lithium niobate near-infrared platform for multiplexing quantum nodes},\
  }\href@noop {} {\bibfield  {journal} {\bibinfo  {journal} {Nature
  communications}\ }\textbf {\bibinfo {volume} {15}},\ \bibinfo {pages} {1}
  (\bibinfo {year} {2024})}\BibitemShut {NoStop}%
\bibitem [{\citenamefont {Cohen}\ \emph {et~al.}(2024)\citenamefont {Cohen},
  \citenamefont {Wu}, \citenamefont {Myilswamy}, \citenamefont {Fatema},
  \citenamefont {Lingaraju},\ and\ \citenamefont {Weiner}}]{cohen2024silicon}%
  \BibitemOpen
  \bibfield  {author} {\bibinfo {author} {\bibfnamefont {L.~M.}\ \bibnamefont
  {Cohen}}, \bibinfo {author} {\bibfnamefont {K.}~\bibnamefont {Wu}}, \bibinfo
  {author} {\bibfnamefont {K.~V.}\ \bibnamefont {Myilswamy}}, \bibinfo {author}
  {\bibfnamefont {S.}~\bibnamefont {Fatema}}, \bibinfo {author} {\bibfnamefont
  {N.~B.}\ \bibnamefont {Lingaraju}},\ and\ \bibinfo {author} {\bibfnamefont
  {A.~M.}\ \bibnamefont {Weiner}},\ }\bibfield  {title} {\bibinfo {title}
  {Silicon photonic microresonator-based high-resolution line-by-line pulse
  shaping},\ }\href@noop {} {\bibfield  {journal} {\bibinfo  {journal} {Nature
  Communications}\ }\textbf {\bibinfo {volume} {15}},\ \bibinfo {pages} {7878}
  (\bibinfo {year} {2024})}\BibitemShut {NoStop}%
\bibitem [{\citenamefont {Wu}\ \emph {et~al.}(2025)\citenamefont {Wu},
  \citenamefont {Cohen}, \citenamefont {Myilswamy}, \citenamefont {Lingaraju},
  \citenamefont {Lu}, \citenamefont {Lukens},\ and\ \citenamefont
  {Weiner}}]{wu2025chip}%
  \BibitemOpen
  \bibfield  {author} {\bibinfo {author} {\bibfnamefont {K.}~\bibnamefont
  {Wu}}, \bibinfo {author} {\bibfnamefont {L.~M.}\ \bibnamefont {Cohen}},
  \bibinfo {author} {\bibfnamefont {K.~V.}\ \bibnamefont {Myilswamy}}, \bibinfo
  {author} {\bibfnamefont {N.~B.}\ \bibnamefont {Lingaraju}}, \bibinfo {author}
  {\bibfnamefont {H.-H.}\ \bibnamefont {Lu}}, \bibinfo {author} {\bibfnamefont
  {J.~M.}\ \bibnamefont {Lukens}},\ and\ \bibinfo {author} {\bibfnamefont
  {A.~M.}\ \bibnamefont {Weiner}},\ }\bibfield  {title} {\bibinfo {title}
  {On-chip pulse shaping of entangled photons},\ }\href@noop {} {\bibfield
  {journal} {\bibinfo  {journal} {Physical Review Research}\ }\textbf {\bibinfo
  {volume} {7}},\ \bibinfo {pages} {033015} (\bibinfo {year}
  {2025})}\BibitemShut {NoStop}%
\bibitem [{\citenamefont {Wang}\ \emph {et~al.}(2023)\citenamefont {Wang},
  \citenamefont {Mere}, \citenamefont {Valdez},\ and\ \citenamefont
  {Mookherjea}}]{wang2023integrated}%
  \BibitemOpen
  \bibfield  {author} {\bibinfo {author} {\bibfnamefont {X.}~\bibnamefont
  {Wang}}, \bibinfo {author} {\bibfnamefont {V.}~\bibnamefont {Mere}}, \bibinfo
  {author} {\bibfnamefont {F.}~\bibnamefont {Valdez}},\ and\ \bibinfo {author}
  {\bibfnamefont {S.}~\bibnamefont {Mookherjea}},\ }\bibfield  {title}
  {\bibinfo {title} {Integrated electro-optic control of biphoton generation
  using hybrid photonics},\ }\href@noop {} {\bibfield  {journal} {\bibinfo
  {journal} {Optica Quantum}\ }\textbf {\bibinfo {volume} {1}},\ \bibinfo
  {pages} {19} (\bibinfo {year} {2023})}\BibitemShut {NoStop}%
\bibitem [{\citenamefont {Oliver}\ \emph {et~al.}(2025)\citenamefont {Oliver},
  \citenamefont {Blau}, \citenamefont {Joshi}, \citenamefont {Ji},
  \citenamefont {Guti{\'e}rrez-J{\'a}uregui}, \citenamefont {Asenjo-Garcia},
  \citenamefont {Lipson},\ and\ \citenamefont {Gaeta}}]{2025_Oliver_NwayBS}%
  \BibitemOpen
  \bibfield  {author} {\bibinfo {author} {\bibfnamefont {R.}~\bibnamefont
  {Oliver}}, \bibinfo {author} {\bibfnamefont {M.}~\bibnamefont {Blau}},
  \bibinfo {author} {\bibfnamefont {C.}~\bibnamefont {Joshi}}, \bibinfo
  {author} {\bibfnamefont {X.}~\bibnamefont {Ji}}, \bibinfo {author}
  {\bibfnamefont {R.}~\bibnamefont {Guti{\'e}rrez-J{\'a}uregui}}, \bibinfo
  {author} {\bibfnamefont {A.}~\bibnamefont {Asenjo-Garcia}}, \bibinfo {author}
  {\bibfnamefont {M.}~\bibnamefont {Lipson}},\ and\ \bibinfo {author}
  {\bibfnamefont {A.~L.}\ \bibnamefont {Gaeta}},\ }\bibfield  {title} {\bibinfo
  {title} {N-way parametric frequency beamsplitter for quantum photonics},\
  }\href@noop {} {\bibfield  {journal} {\bibinfo  {journal} {Physical Review
  Research}\ }\textbf {\bibinfo {volume} {7}},\ \bibinfo {pages} {023108}
  (\bibinfo {year} {2025})}\BibitemShut {NoStop}%
\bibitem [{\citenamefont {Folge}\ \emph {et~al.}(2024)\citenamefont {Folge},
  \citenamefont {Stefszky}, \citenamefont {Brecht},\ and\ \citenamefont
  {Silberhorn}}]{folge2024framework}%
  \BibitemOpen
  \bibfield  {author} {\bibinfo {author} {\bibfnamefont {P.}~\bibnamefont
  {Folge}}, \bibinfo {author} {\bibfnamefont {M.}~\bibnamefont {Stefszky}},
  \bibinfo {author} {\bibfnamefont {B.}~\bibnamefont {Brecht}},\ and\ \bibinfo
  {author} {\bibfnamefont {C.}~\bibnamefont {Silberhorn}},\ }\bibfield  {title}
  {\bibinfo {title} {A framework for fully programmable frequency-encoded
  quantum networks harnessing multioutput quantum pulse gates},\ }\href@noop {}
  {\bibfield  {journal} {\bibinfo  {journal} {PRX Quantum}\ }\textbf {\bibinfo
  {volume} {5}},\ \bibinfo {pages} {040329} (\bibinfo {year}
  {2024})}\BibitemShut {NoStop}%
\bibitem [{\citenamefont {Boeuf}\ \emph {et~al.}(2021)\citenamefont {Boeuf},
  \citenamefont {Barrera}, \citenamefont {Fincato}, \citenamefont {Tang},
  \citenamefont {Guerber}, \citenamefont {Monfray}, \citenamefont {Ohno},
  \citenamefont {Fowler}, \citenamefont {Charlet}, \citenamefont {Gianini},
  \citenamefont {Simbula}, \citenamefont {Maggi}, \citenamefont {Shaw},
  \citenamefont {Toprasertpong}, \citenamefont {Takagi},\ and\ \citenamefont
  {Takenaka}}]{2021_Boeuf_ST}%
  \BibitemOpen
  \bibfield  {author} {\bibinfo {author} {\bibfnamefont {F.}~\bibnamefont
  {Boeuf}}, \bibinfo {author} {\bibfnamefont {C.}~\bibnamefont {Barrera}},
  \bibinfo {author} {\bibfnamefont {A.}~\bibnamefont {Fincato}}, \bibinfo
  {author} {\bibfnamefont {H.}~\bibnamefont {Tang}}, \bibinfo {author}
  {\bibfnamefont {S.}~\bibnamefont {Guerber}}, \bibinfo {author} {\bibfnamefont
  {S.}~\bibnamefont {Monfray}}, \bibinfo {author} {\bibfnamefont
  {S.}~\bibnamefont {Ohno}}, \bibinfo {author} {\bibfnamefont {D.}~\bibnamefont
  {Fowler}}, \bibinfo {author} {\bibfnamefont {I.}~\bibnamefont {Charlet}},
  \bibinfo {author} {\bibfnamefont {L.}~\bibnamefont {Gianini}}, \bibinfo
  {author} {\bibfnamefont {A.}~\bibnamefont {Simbula}}, \bibinfo {author}
  {\bibfnamefont {L.}~\bibnamefont {Maggi}}, \bibinfo {author} {\bibfnamefont
  {M.}~\bibnamefont {Shaw}}, \bibinfo {author} {\bibfnamefont {K.}~\bibnamefont
  {Toprasertpong}}, \bibinfo {author} {\bibfnamefont {S.}~\bibnamefont
  {Takagi}},\ and\ \bibinfo {author} {\bibfnamefont {M.}~\bibnamefont
  {Takenaka}},\ }\bibfield  {title} {\bibinfo {title} {Silicon photonics beyond
  optical interconnects},\ }in\ \href
  {https://doi.org/10.1109/IEDM19574.2021.9720504} {\emph {\bibinfo {booktitle}
  {2021 IEEE International Electron Devices Meeting (IEDM)}}}\ (\bibinfo {year}
  {2021})\ pp.\ \bibinfo {pages} {29.2.1--29.2.4}\BibitemShut {NoStop}%
\bibitem [{\citenamefont {Sharma}\ \emph {et~al.}(2025)\citenamefont {Sharma},
  \citenamefont {Brandm{\"u}ller}, \citenamefont {B{\"u}tow}, \citenamefont
  {Eismann},\ and\ \citenamefont {Banzer}}]{sharma2025universal}%
  \BibitemOpen
  \bibfield  {author} {\bibinfo {author} {\bibfnamefont {V.}~\bibnamefont
  {Sharma}}, \bibinfo {author} {\bibfnamefont {D.}~\bibnamefont
  {Brandm{\"u}ller}}, \bibinfo {author} {\bibfnamefont {J.}~\bibnamefont
  {B{\"u}tow}}, \bibinfo {author} {\bibfnamefont {J.~S.}\ \bibnamefont
  {Eismann}},\ and\ \bibinfo {author} {\bibfnamefont {P.}~\bibnamefont
  {Banzer}},\ }\bibfield  {title} {\bibinfo {title} {Universal photonic
  processor for spatial mode decomposition},\ }\href@noop {} {\bibfield
  {journal} {\bibinfo  {journal} {Nature Communications}\ }\textbf {\bibinfo
  {volume} {16}},\ \bibinfo {pages} {7982} (\bibinfo {year}
  {2025})}\BibitemShut {NoStop}%
\bibitem [{\citenamefont {Lu}\ \emph {et~al.}(2018{\natexlab{b}})\citenamefont
  {Lu}, \citenamefont {Lukens}, \citenamefont {Peters}, \citenamefont
  {Williams}, \citenamefont {Weiner},\ and\ \citenamefont
  {Lougovski}}]{2018_Lu_FB_Q}%
  \BibitemOpen
  \bibfield  {author} {\bibinfo {author} {\bibfnamefont {H.-H.}\ \bibnamefont
  {Lu}}, \bibinfo {author} {\bibfnamefont {J.~M.}\ \bibnamefont {Lukens}},
  \bibinfo {author} {\bibfnamefont {N.~A.}\ \bibnamefont {Peters}}, \bibinfo
  {author} {\bibfnamefont {B.~P.}\ \bibnamefont {Williams}}, \bibinfo {author}
  {\bibfnamefont {A.~M.}\ \bibnamefont {Weiner}},\ and\ \bibinfo {author}
  {\bibfnamefont {P.}~\bibnamefont {Lougovski}},\ }\bibfield  {title} {\bibinfo
  {title} {Quantum interference and correlation control of frequency-bin
  qubits},\ }\href@noop {} {\bibfield  {journal} {\bibinfo  {journal} {Optica}\
  }\textbf {\bibinfo {volume} {5}},\ \bibinfo {pages} {1455} (\bibinfo {year}
  {2018}{\natexlab{b}})}\BibitemShut {NoStop}%
\bibitem [{\citenamefont {Lu}\ \emph {et~al.}(2020)\citenamefont {Lu},
  \citenamefont {Simmerman}, \citenamefont {Lougovski}, \citenamefont
  {Weiner},\ and\ \citenamefont {Lukens}}]{2020_Lu_fully}%
  \BibitemOpen
  \bibfield  {author} {\bibinfo {author} {\bibfnamefont {H.-H.}\ \bibnamefont
  {Lu}}, \bibinfo {author} {\bibfnamefont {E.~M.}\ \bibnamefont {Simmerman}},
  \bibinfo {author} {\bibfnamefont {P.}~\bibnamefont {Lougovski}}, \bibinfo
  {author} {\bibfnamefont {A.~M.}\ \bibnamefont {Weiner}},\ and\ \bibinfo
  {author} {\bibfnamefont {J.~M.}\ \bibnamefont {Lukens}},\ }\bibfield  {title}
  {\bibinfo {title} {Fully arbitrary control of frequency-bin qubits},\
  }\href@noop {} {\bibfield  {journal} {\bibinfo  {journal} {Physical Review
  Letters}\ }\textbf {\bibinfo {volume} {125}},\ \bibinfo {pages} {120503}
  (\bibinfo {year} {2020})}\BibitemShut {NoStop}%
\bibitem [{\citenamefont {Rahim}\ \emph {et~al.}(2021)\citenamefont {Rahim},
  \citenamefont {Hermans}, \citenamefont {Wohlfeil}, \citenamefont {Petousi},
  \citenamefont {Kuyken}, \citenamefont {Van~Thourhout},\ and\ \citenamefont
  {Baets}}]{rahim2021taking}%
  \BibitemOpen
  \bibfield  {author} {\bibinfo {author} {\bibfnamefont {A.}~\bibnamefont
  {Rahim}}, \bibinfo {author} {\bibfnamefont {A.}~\bibnamefont {Hermans}},
  \bibinfo {author} {\bibfnamefont {B.}~\bibnamefont {Wohlfeil}}, \bibinfo
  {author} {\bibfnamefont {D.}~\bibnamefont {Petousi}}, \bibinfo {author}
  {\bibfnamefont {B.}~\bibnamefont {Kuyken}}, \bibinfo {author} {\bibfnamefont
  {D.}~\bibnamefont {Van~Thourhout}},\ and\ \bibinfo {author} {\bibfnamefont
  {R.}~\bibnamefont {Baets}},\ }\bibfield  {title} {\bibinfo {title} {Taking
  silicon photonics modulators to a higher performance level: state-of-the-art
  and a review of new technologies},\ }\href@noop {} {\bibfield  {journal}
  {\bibinfo  {journal} {Advanced Photonics}\ }\textbf {\bibinfo {volume} {3}},\
  \bibinfo {pages} {024003} (\bibinfo {year} {2021})}\BibitemShut {NoStop}%
\bibitem [{\citenamefont {Taghavi}\ \emph {et~al.}(2022)\citenamefont
  {Taghavi}, \citenamefont {Moridsadat}, \citenamefont {Tofini}, \citenamefont
  {Raza}, \citenamefont {Jaeger}, \citenamefont {Chrostowski}, \citenamefont
  {Shastri},\ and\ \citenamefont {Shekhar}}]{taghavi2022polymer}%
  \BibitemOpen
  \bibfield  {author} {\bibinfo {author} {\bibfnamefont {I.}~\bibnamefont
  {Taghavi}}, \bibinfo {author} {\bibfnamefont {M.}~\bibnamefont {Moridsadat}},
  \bibinfo {author} {\bibfnamefont {A.}~\bibnamefont {Tofini}}, \bibinfo
  {author} {\bibfnamefont {S.}~\bibnamefont {Raza}}, \bibinfo {author}
  {\bibfnamefont {N.~A.}\ \bibnamefont {Jaeger}}, \bibinfo {author}
  {\bibfnamefont {L.}~\bibnamefont {Chrostowski}}, \bibinfo {author}
  {\bibfnamefont {B.~J.}\ \bibnamefont {Shastri}},\ and\ \bibinfo {author}
  {\bibfnamefont {S.}~\bibnamefont {Shekhar}},\ }\bibfield  {title} {\bibinfo
  {title} {Polymer modulators in silicon photonics: review and projections},\
  }\href@noop {} {\bibfield  {journal} {\bibinfo  {journal} {Nanophotonics}\
  }\textbf {\bibinfo {volume} {11}},\ \bibinfo {pages} {3855} (\bibinfo {year}
  {2022})}\BibitemShut {NoStop}%
\bibitem [{\citenamefont {Han}\ \emph {et~al.}(2025{\natexlab{b}})\citenamefont
  {Han}, \citenamefont {Ruan},\ and\ \citenamefont
  {Xiang}}]{han2025heterogeneously}%
  \BibitemOpen
  \bibfield  {author} {\bibinfo {author} {\bibfnamefont {H.}~\bibnamefont
  {Han}}, \bibinfo {author} {\bibfnamefont {S.}~\bibnamefont {Ruan}},\ and\
  \bibinfo {author} {\bibfnamefont {B.}~\bibnamefont {Xiang}},\ }\bibfield
  {title} {\bibinfo {title} {Heterogeneously integrated photonics based on thin
  film lithium niobate platform},\ }\href@noop {} {\bibfield  {journal}
  {\bibinfo  {journal} {Laser \& Photonics Reviews}\ }\textbf {\bibinfo
  {volume} {19}},\ \bibinfo {pages} {2400649} (\bibinfo {year}
  {2025}{\natexlab{b}})}\BibitemShut {NoStop}%
\bibitem [{\citenamefont {Buddhiraju}\ \emph {et~al.}(2021)\citenamefont
  {Buddhiraju}, \citenamefont {Dutt}, \citenamefont {Minkov}, \citenamefont
  {Williamson},\ and\ \citenamefont {Fan}}]{buddhiraju2021arbitrary}%
  \BibitemOpen
  \bibfield  {author} {\bibinfo {author} {\bibfnamefont {S.}~\bibnamefont
  {Buddhiraju}}, \bibinfo {author} {\bibfnamefont {A.}~\bibnamefont {Dutt}},
  \bibinfo {author} {\bibfnamefont {M.}~\bibnamefont {Minkov}}, \bibinfo
  {author} {\bibfnamefont {I.~A.}\ \bibnamefont {Williamson}},\ and\ \bibinfo
  {author} {\bibfnamefont {S.}~\bibnamefont {Fan}},\ }\bibfield  {title}
  {\bibinfo {title} {Arbitrary linear transformations for photons in the
  frequency synthetic dimension},\ }\href@noop {} {\bibfield  {journal}
  {\bibinfo  {journal} {Nature communications}\ }\textbf {\bibinfo {volume}
  {12}},\ \bibinfo {pages} {2401} (\bibinfo {year} {2021})}\BibitemShut
  {NoStop}%
\bibitem [{\citenamefont {Hu}\ \emph {et~al.}(2021)\citenamefont {Hu},
  \citenamefont {Yu}, \citenamefont {Zhu}, \citenamefont {Sinclair},
  \citenamefont {Shams-Ansari}, \citenamefont {Shao}, \citenamefont
  {Holzgrafe}, \citenamefont {Puma}, \citenamefont {Zhang},\ and\ \citenamefont
  {Lon{\v{c}}ar}}]{hu2021chip}%
  \BibitemOpen
  \bibfield  {author} {\bibinfo {author} {\bibfnamefont {Y.}~\bibnamefont
  {Hu}}, \bibinfo {author} {\bibfnamefont {M.}~\bibnamefont {Yu}}, \bibinfo
  {author} {\bibfnamefont {D.}~\bibnamefont {Zhu}}, \bibinfo {author}
  {\bibfnamefont {N.}~\bibnamefont {Sinclair}}, \bibinfo {author}
  {\bibfnamefont {A.}~\bibnamefont {Shams-Ansari}}, \bibinfo {author}
  {\bibfnamefont {L.}~\bibnamefont {Shao}}, \bibinfo {author} {\bibfnamefont
  {J.}~\bibnamefont {Holzgrafe}}, \bibinfo {author} {\bibfnamefont
  {E.}~\bibnamefont {Puma}}, \bibinfo {author} {\bibfnamefont {M.}~\bibnamefont
  {Zhang}},\ and\ \bibinfo {author} {\bibfnamefont {M.}~\bibnamefont
  {Lon{\v{c}}ar}},\ }\bibfield  {title} {\bibinfo {title} {On-chip
  electro-optic frequency shifters and beam splitters},\ }\href@noop {}
  {\bibfield  {journal} {\bibinfo  {journal} {Nature}\ }\textbf {\bibinfo
  {volume} {599}},\ \bibinfo {pages} {587} (\bibinfo {year}
  {2021})}\BibitemShut {NoStop}%
\bibitem [{\citenamefont {Reck}\ \emph {et~al.}(1994)\citenamefont {Reck},
  \citenamefont {Zeilinger}, \citenamefont {Bernstein},\ and\ \citenamefont
  {Bertani}}]{reck1994experimental}%
  \BibitemOpen
  \bibfield  {author} {\bibinfo {author} {\bibfnamefont {M.}~\bibnamefont
  {Reck}}, \bibinfo {author} {\bibfnamefont {A.}~\bibnamefont {Zeilinger}},
  \bibinfo {author} {\bibfnamefont {H.~J.}\ \bibnamefont {Bernstein}},\ and\
  \bibinfo {author} {\bibfnamefont {P.}~\bibnamefont {Bertani}},\ }\bibfield
  {title} {\bibinfo {title} {Experimental realization of any discrete unitary
  operator},\ }\href@noop {} {\bibfield  {journal} {\bibinfo  {journal}
  {Physical review letters}\ }\textbf {\bibinfo {volume} {73}},\ \bibinfo
  {pages} {58} (\bibinfo {year} {1994})}\BibitemShut {NoStop}%
\bibitem [{\citenamefont {Karnieli}\ \emph {et~al.}(2025)\citenamefont
  {Karnieli}, \citenamefont {Mor}, \citenamefont {Roques-Carmes}, \citenamefont
  {Lustig}, \citenamefont {Sloan}, \citenamefont {Vu{\v{c}}kovi{\'c}},
  \citenamefont {Miller},\ and\ \citenamefont {Fan}}]{karnieli2025variational}%
  \BibitemOpen
  \bibfield  {author} {\bibinfo {author} {\bibfnamefont {A.}~\bibnamefont
  {Karnieli}}, \bibinfo {author} {\bibfnamefont {P.-A.}\ \bibnamefont {Mor}},
  \bibinfo {author} {\bibfnamefont {C.}~\bibnamefont {Roques-Carmes}}, \bibinfo
  {author} {\bibfnamefont {E.}~\bibnamefont {Lustig}}, \bibinfo {author}
  {\bibfnamefont {J.}~\bibnamefont {Sloan}}, \bibinfo {author} {\bibfnamefont
  {J.}~\bibnamefont {Vu{\v{c}}kovi{\'c}}}, \bibinfo {author} {\bibfnamefont
  {D.~A.}\ \bibnamefont {Miller}},\ and\ \bibinfo {author} {\bibfnamefont
  {S.}~\bibnamefont {Fan}},\ }\bibfield  {title} {\bibinfo {title} {Variational
  processing of multimode squeezed light},\ }\href@noop {} {\bibfield
  {journal} {\bibinfo  {journal} {arXiv preprint arXiv:2509.16753}\ } (\bibinfo
  {year} {2025})}\BibitemShut {NoStop}%
\bibitem [{\citenamefont {Bunandar}\ \emph {et~al.}(2018)\citenamefont
  {Bunandar}, \citenamefont {Lentine}, \citenamefont {Lee}, \citenamefont
  {Cai}, \citenamefont {Long}, \citenamefont {Boynton}, \citenamefont
  {Martinez}, \citenamefont {DeRose}, \citenamefont {Chen}, \citenamefont
  {Grein} \emph {et~al.}}]{bunandar2018metropolitan}%
  \BibitemOpen
  \bibfield  {author} {\bibinfo {author} {\bibfnamefont {D.}~\bibnamefont
  {Bunandar}}, \bibinfo {author} {\bibfnamefont {A.}~\bibnamefont {Lentine}},
  \bibinfo {author} {\bibfnamefont {C.}~\bibnamefont {Lee}}, \bibinfo {author}
  {\bibfnamefont {H.}~\bibnamefont {Cai}}, \bibinfo {author} {\bibfnamefont
  {C.~M.}\ \bibnamefont {Long}}, \bibinfo {author} {\bibfnamefont
  {N.}~\bibnamefont {Boynton}}, \bibinfo {author} {\bibfnamefont
  {N.}~\bibnamefont {Martinez}}, \bibinfo {author} {\bibfnamefont
  {C.}~\bibnamefont {DeRose}}, \bibinfo {author} {\bibfnamefont
  {C.}~\bibnamefont {Chen}}, \bibinfo {author} {\bibfnamefont {M.}~\bibnamefont
  {Grein}}, \emph {et~al.},\ }\bibfield  {title} {\bibinfo {title}
  {Metropolitan quantum key distribution with silicon photonics},\ }\href@noop
  {} {\bibfield  {journal} {\bibinfo  {journal} {Physical Review X}\ }\textbf
  {\bibinfo {volume} {8}},\ \bibinfo {pages} {021009} (\bibinfo {year}
  {2018})}\BibitemShut {NoStop}%
\bibitem [{\citenamefont {Sibson}\ \emph {et~al.}(2017)\citenamefont {Sibson},
  \citenamefont {Kennard}, \citenamefont {Stanisic}, \citenamefont {Erven},
  \citenamefont {O’Brien},\ and\ \citenamefont
  {Thompson}}]{sibson2017integrated}%
  \BibitemOpen
  \bibfield  {author} {\bibinfo {author} {\bibfnamefont {P.}~\bibnamefont
  {Sibson}}, \bibinfo {author} {\bibfnamefont {J.~E.}\ \bibnamefont {Kennard}},
  \bibinfo {author} {\bibfnamefont {S.}~\bibnamefont {Stanisic}}, \bibinfo
  {author} {\bibfnamefont {C.}~\bibnamefont {Erven}}, \bibinfo {author}
  {\bibfnamefont {J.~L.}\ \bibnamefont {O’Brien}},\ and\ \bibinfo {author}
  {\bibfnamefont {M.~G.}\ \bibnamefont {Thompson}},\ }\bibfield  {title}
  {\bibinfo {title} {Integrated silicon photonics for high-speed quantum key
  distribution},\ }\href@noop {} {\bibfield  {journal} {\bibinfo  {journal}
  {Optica}\ }\textbf {\bibinfo {volume} {4}},\ \bibinfo {pages} {172} (\bibinfo
  {year} {2017})}\BibitemShut {NoStop}%
\bibitem [{\citenamefont {Wei}\ \emph {et~al.}(2020)\citenamefont {Wei},
  \citenamefont {Li}, \citenamefont {Tan}, \citenamefont {Li}, \citenamefont
  {Min}, \citenamefont {Zhang}, \citenamefont {Li}, \citenamefont {You},
  \citenamefont {Wang}, \citenamefont {Jiang} \emph {et~al.}}]{wei2020high}%
  \BibitemOpen
  \bibfield  {author} {\bibinfo {author} {\bibfnamefont {K.}~\bibnamefont
  {Wei}}, \bibinfo {author} {\bibfnamefont {W.}~\bibnamefont {Li}}, \bibinfo
  {author} {\bibfnamefont {H.}~\bibnamefont {Tan}}, \bibinfo {author}
  {\bibfnamefont {Y.}~\bibnamefont {Li}}, \bibinfo {author} {\bibfnamefont
  {H.}~\bibnamefont {Min}}, \bibinfo {author} {\bibfnamefont {W.-J.}\
  \bibnamefont {Zhang}}, \bibinfo {author} {\bibfnamefont {H.}~\bibnamefont
  {Li}}, \bibinfo {author} {\bibfnamefont {L.}~\bibnamefont {You}}, \bibinfo
  {author} {\bibfnamefont {Z.}~\bibnamefont {Wang}}, \bibinfo {author}
  {\bibfnamefont {X.}~\bibnamefont {Jiang}}, \emph {et~al.},\ }\bibfield
  {title} {\bibinfo {title} {High-speed measurement-device-independent quantum
  key distribution with integrated silicon photonics},\ }\href@noop {}
  {\bibfield  {journal} {\bibinfo  {journal} {Physical Review X}\ }\textbf
  {\bibinfo {volume} {10}},\ \bibinfo {pages} {031030} (\bibinfo {year}
  {2020})}\BibitemShut {NoStop}%
\bibitem [{\citenamefont {Boeuf}\ \emph {et~al.}(2019)\citenamefont {Boeuf},
  \citenamefont {Fincato}, \citenamefont {Maggi}, \citenamefont {Carpentier},
  \citenamefont {Le~Maitre}, \citenamefont {Shaw}, \citenamefont {Cremer},
  \citenamefont {Vulliet}, \citenamefont {Baudot}, \citenamefont {Monfray}
  \emph {et~al.}}]{2019_Boeuf_ST}%
  \BibitemOpen
  \bibfield  {author} {\bibinfo {author} {\bibfnamefont {F.}~\bibnamefont
  {Boeuf}}, \bibinfo {author} {\bibfnamefont {A.}~\bibnamefont {Fincato}},
  \bibinfo {author} {\bibfnamefont {L.}~\bibnamefont {Maggi}}, \bibinfo
  {author} {\bibfnamefont {J.}~\bibnamefont {Carpentier}}, \bibinfo {author}
  {\bibfnamefont {P.}~\bibnamefont {Le~Maitre}}, \bibinfo {author}
  {\bibfnamefont {M.}~\bibnamefont {Shaw}}, \bibinfo {author} {\bibfnamefont
  {S.}~\bibnamefont {Cremer}}, \bibinfo {author} {\bibfnamefont
  {N.}~\bibnamefont {Vulliet}}, \bibinfo {author} {\bibfnamefont
  {C.}~\bibnamefont {Baudot}}, \bibinfo {author} {\bibfnamefont
  {S.}~\bibnamefont {Monfray}}, \emph {et~al.},\ }\bibfield  {title} {\bibinfo
  {title} {A silicon photonics technology for 400 gbit/s applications},\ }in\
  \href@noop {} {\emph {\bibinfo {booktitle} {2019 IEEE international electron
  devices meeting (IEDM)}}}\ (\bibinfo {organization} {IEEE},\ \bibinfo {year}
  {2019})\ pp.\ \bibinfo {pages} {33--1}\BibitemShut {NoStop}%
\bibitem [{\citenamefont {Monfray}\ \emph {et~al.}(2020)\citenamefont
  {Monfray}, \citenamefont {Cremer}, \citenamefont {Vulliet}, \citenamefont
  {Dubois}, \citenamefont {Domengie}, \citenamefont {Jan}, \citenamefont
  {Baille},\ and\ \citenamefont {Boeuf}}]{2020_Boeuf_STopt}%
  \BibitemOpen
  \bibfield  {author} {\bibinfo {author} {\bibfnamefont {S.}~\bibnamefont
  {Monfray}}, \bibinfo {author} {\bibfnamefont {S.}~\bibnamefont {Cremer}},
  \bibinfo {author} {\bibfnamefont {N.}~\bibnamefont {Vulliet}}, \bibinfo
  {author} {\bibfnamefont {E.}~\bibnamefont {Dubois}}, \bibinfo {author}
  {\bibfnamefont {F.}~\bibnamefont {Domengie}}, \bibinfo {author}
  {\bibfnamefont {S.}~\bibnamefont {Jan}}, \bibinfo {author} {\bibfnamefont
  {F.}~\bibnamefont {Baille}},\ and\ \bibinfo {author} {\bibfnamefont
  {F.}~\bibnamefont {Boeuf}},\ }\bibfield  {title} {\bibinfo {title}
  {Optimization of deep rib high speed phase modulators on 300mm industrial
  si-photonics platform},\ }in\ \href@noop {} {\emph {\bibinfo {booktitle}
  {Integrated Photonics Platforms: Fundamental Research, Manufacturing and
  Applications}}},\ Vol.\ \bibinfo {volume} {11364}\ (\bibinfo {organization}
  {SPIE},\ \bibinfo {year} {2020})\ pp.\ \bibinfo {pages} {7--14}\BibitemShut
  {NoStop}%
\end{thebibliography}

%

\end{document}